\documentclass[aps,pre,superscriptaddress,twocolumn]{revtex4-2}

\usepackage{aligned-overset}
\usepackage{amsmath}
\usepackage{amssymb}
\usepackage{booktabs}
\usepackage{array}
\usepackage{geometry}
\usepackage{graphicx}
\usepackage{xspace}
\usepackage{xcolor}
\usepackage[english]{babel}
\usepackage{hyperref}

\hypersetup{
  bookmarksopen=false,
  bookmarksnumbered=true,
  colorlinks=true,
  linkcolor=black,
  citecolor=black,
  urlcolor=black,
  pdftitle={SCEPTR},
  pdfauthor={Yuta Nagano et al.},
}

\newcommand{\tcr}{\sigma}

\newcommand{\reals}{\mathbb{R}}
\newcommand{\integers}{\mathbb{Z}}
\newcommand{\Loss}[1]{\mathcal{L}_\textrm{#1}}
\newcommand{\p}[1]{p_\textrm{#1}}
\newcommand{\uhsphere}{\mathcal{S}}
\newcommand{\cls}{\texttt{<cls>}\xspace}
\newcommand{\mask}{\texttt{<mask>}\xspace}
\newcommand{\SCEPTRFull}{\underline{S}imple \underline{C}ontrastive \underline{E}mbedding of the \underline{P}rimary sequence of \underline{T} cell \underline{R}eceptors\xspace}
\DeclareMathOperator*{\E}{\mathbb{E}}

\begin{document}

\title{Contrastive learning of T cell receptor representations}

\author{Yuta Nagano}
\affiliation{Division of Medicine, University College London}
\affiliation{Division of Infection and Immunity, University College London}

\author{Andrew Pyo}
\affiliation{Center for the Physics of Biological Function, Princeton University}

\author{Martina Milighetti}
\affiliation{Division of Infection and Immunity, University College London}
\affiliation{Cancer Institute, University College London}

\author{James Henderson}
\affiliation{Division of Infection and Immunity, University College London}
\affiliation{Institute for the Physics of Living Systems, University College London}

\author{John Shawe-Taylor}
\affiliation{Department of Computer Science, University College London}

\author{Benny Chain}
\thanks{Joint last authors}
\affiliation{Division of Infection and Immunity, University College London}
\affiliation{Department of Computer Science, University College London}

\author{Andreas Tiffeau-Mayer}
\thanks{Joint last authors}
\affiliation{Division of Infection and Immunity, University College London}
\affiliation{Institute for the Physics of Living Systems, University College London}

\begin{abstract}

Computational prediction of the interaction of T cell receptors (TCRs) and their ligands is a grand challenge in immunology.
Despite advances in high-throughput assays, specificity-labelled TCR data remains sparse.
In other domains, the pre-training of language models on unlabelled data has been successfully used to address data bottlenecks.
However, it is unclear how to best pre-train protein language models for TCR specificity prediction.
Here we introduce a TCR language model called SCEPTR (\SCEPTRFull), capable of data-efficient transfer learning.
Through our model, we introduce a novel pre-training strategy combining autocontrastive learning and masked-language modelling, which enables SCEPTR to achieve its state-of-the-art performance.
In contrast, existing protein language models and a variant of SCEPTR pre-trained without autocontrastive learning are outperformed by sequence alignment-based methods.
We anticipate that contrastive learning will be a useful paradigm to decode the rules of TCR specificity.

\end{abstract}

\maketitle

Antigen-specific T cells play important protective and pathogenic roles in human disease \cite{chi2024principles}.
The recognition of peptides presented on major histocompatibility complexes (pMHCs) by $\alpha\beta$ T cell receptors (TCRs) determines the specificity of cellular immune responses~\cite{Davis1988CellReceptor}.
Hyperdiverse $\alpha\beta$TCRs are generated during T cell development in the thymus by genetic recombination of germline-encoded V, D (for TCR$\beta$) and J gene segments with additional diversification by trimming of the germline and insertions of non-template nucleotides at gene segment junctions.

A major goal of systems immunology is to uncover the rules governing which TCRs interact with which pMHCs \cite{hudson_can_2023}.
Advances in high-throughput functional assays of TCR specificity \cite{dash_quantifiable_2017,dobson2022antigen,joglekar2019t} have made the use of machine learning a promising prospect to discover such rules.

The most direct approach for applying machine learning to TCR specificity prediction has been to train pMHC-specific models that take an arbitrary TCR and predict binding~\cite{gielis_detection_2019, fischer_predicting_2020, jokinen_predicting_2021,montemurro_nettcr-20_2021,wu_tcr-bert_2021,croce2024deep}.
More ambitiously, model architectures have been proposed that can in principle generalise predictions to arbitrary pMHCs as well~\cite{weber_titan_2021, jiang_teinet_2023, lu_deep_2021, lin_rapid_2021, springer_contribution_2021, cai_atm-tcr_2022, moris_current_2021, pham_epitcr_2023, gao_pan-peptide_2023, kwee_stapler_2023, meynard-piganeau_tulip_2024, GoldnerKabeli2024SelfsupervisedLearning}.
Independent benchmarking studies have shown that both approaches are effective for predicting TCR binders against pMHCs for which many TCRs have been experimentally determined~\cite{nielsen_lessons_2024}, but generalisation to pMHCs not seen during training has largely remained elusive~\cite{grazioli2022tcr} and prediction accuracy is limited for pMHCs with few known binders \cite{deng2023performance}.
This severely limits the utility of current predictive tools given that to date, only $\sim 10^3$ of the $> 10^{15}$ possible pMHCs are annotated with any TCRs in VDJdb \cite{bagaev_vdjdb_2020}, and given that for $>95\%$ of them less than 100 specific TCRs are known.

Meanwhile, there is abundant unlabelled TCR sequence data that may be exploited for unsupervised representation learning.
A TCR representation model that compactly captures important features would provide embeddings useful for data-efficient training of downstream specificity predictors.

In natural language processing (NLP), unsupervised pre-trained transformers have demonstrated capacity for transfer learning to diverse downstream tasks \cite{vaswani_attention_2017,devlin_bert_2019,brown2020language}.
This has spurred substantial work applying transformers to protein analysis.
Protein language models (PLMs) such as those of the ESM~\cite{rives2021biological,lin_evolutionary-scale_2023} and ProtTrans~\cite{elnaggar_prottrans_2022} families have been successfully used in structure-prediction pipelines and for protein property prediction \cite{elnaggar_ankh_2023,wu_proteinclip_2024,li_feature_2024}.
PLMs have also been applied to TCR-pMHC interaction prediction~\cite{wu_tcr-bert_2021, kwee_stapler_2023, meynard-piganeau_tulip_2024, GoldnerKabeli2024SelfsupervisedLearning}, and the related problem of antibody-antigen interaction prediction \cite{wang2024language,barton_generative_2024}.
However, there has been limited systematic testing of how competitive PLM embeddings are in the \emph{few-shot} setting typical for most ligands -- that is, where only few labelled data points are available for transfer learning.

To address this question, we benchmarked existing PLMs on a standardised few-shot specificity prediction task, and surprisingly found that they are inferior to state-of-the-art sequence alignment-based methods.
This motivated us to develop SCEPTR (\SCEPTRFull), a novel TCR PLM which closes this gap.
Our key innovation is a pre-training strategy involving an autocontrastive learning procedure adapted for $\alpha\beta$TCRs, which we show is the primary driver behind SCEPTR's improved performance.

\begin{figure*}
  \includegraphics{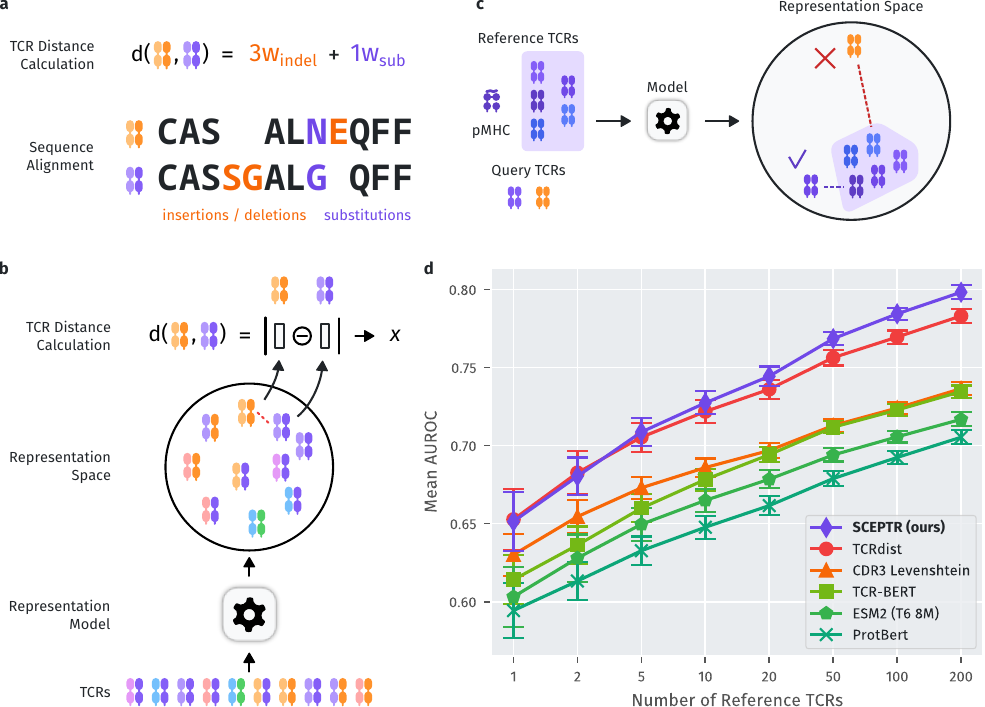}
  \caption{
    \textbf{Benchmarking TCR language models against sequence alignment-based approaches on few-shot TCR specificity prediction.}
    \textbf{a)} TCR similarity can be quantified using sequence-alignment by taking a (weighted) count of how many sequence edits turn one TCR into another.
    \textbf{b)} Learned sequence representations allow alignment-free sequence comparisons based on distances in the embedding feature space.
    \textbf{c)} Sketch of our standardized benchmarking approach to allow side-by-side comparison of sequence-alignment and embedding methods. Using a reference set of known TCR binders to a pMHC of interest, we propose nearest neighbour prediction as a task for unbiased comparison of the quality of embeddings for specificity prediction.
    \textbf{d)} Performance of six different models on TCR specificity prediction as a function of the number of reference TCRs.
    Specificity predictions were made by the nearest neighbour method sketched in \textbf{c} against six different pMHCs and performance is reported as the AUROC averaged across the pMHCs. 
    The error bars represent standard deviations of model AUROCs relative to the average across all models within a data split.
  }
  \label{fig:benchmarking_overview}
\end{figure*}

\section{Results}

\subsection{Benchmarking PLM embeddings on TCR specificity prediction}
\label{sec:benchmarking}

Given the scarcity of specificity-labelled TCR data, it is of practical importance to evaluate model performance where access to such data is limited.
Therefore, we set up a benchmarking framework focused on few-shot TCR specificity prediction.

To conduct our benchmark, we curated a set of specificity-labelled $\alpha\beta$TCR data from VDJdb~\cite{bagaev_vdjdb_2020}.
We only included human TCRs with full $\alpha$ and $\beta$ chain information, and excluded data from an early 10x Genomics whitepaper~\cite{10x_genomics_new_2020}, as there are known issues with data reliability in this study~\cite{zhang_framework_2021, montemurro_benchmarking_2023}.
This left us with a total of 7168 $\alpha\beta$TCRs annotated to 864 pMHCs.
Of these, we used the six pMHCs with greater than 300 distinct binder TCRs for our benchmarking task.

We created a benchmarking task that allowed us to directly compare sequence alignment-based distance metrics such as the state-of-the-art  TCRdist~\cite{dash_quantifiable_2017,mayer2021tcr} (Fig.~\ref{fig:benchmarking_overview}a) to distances in PLM embedding spaces (Fig.~\ref{fig:benchmarking_overview}b).
For each pMHC, we tested models on their ability to distinguish binder TCRs from non-binders using embedding distances between a query TCR and its closest neighbour within a reference set (Fig.~\ref{fig:benchmarking_overview}c).
We call this \emph{nearest neighbour prediction}.
This framework is simple and attractive for benchmarking models in the few-shot regime, since it remains well defined for as few as a single reference TCR and does not require model specific fine-tuning.

We conducted multiple benchmarks for each pMHC, varying the number of its cognate TCRs used as the reference set.
In each case, we combined the remaining TCRs for the target with the rest of the filtered VDJdb dataset (including TCRs annotated to pMHCs other than the six target pMHCs) to create a test set (see methods~\ref{sec:methods_benchmarking}).
By studying how performance depends on the size of the reference set, we are effectively probing representation alignment with TCR co-specificity prediction at different scales.

We benchmarked six models: two alignment-based TCR metrics (CDR3 Levenshtein distance and TCRdist~\cite{dash_quantifiable_2017}), two general-purpose PLMs (ProtBert~\cite{elnaggar_prottrans_2022} and ESM2~\cite{lin_evolutionary-scale_2023}), and two TCR domain-specific language models (TCR-BERT~\cite{wu_tcr-bert_2021} and our own model SCEPTR).
We report performance using the area under the receiver operator characteristic (AUROC) averaged over the tested pMHCs.

To our surprise, we found that TCR-BERT, ESM2, and ProtBert all fail to outperform the baseline sequence alignment method (CDR3 Levenshtein) and are significantly inferior to TCRdist (Figs.~\ref{fig:benchmarking_overview}d / \ref{fig:individual_rocs}).
A repeat of the benchmarking with a broader set of epitopes obtained by including post-processed 10x Genomics whitepaper data~\cite{montemurro_benchmarking_2023} recapitulated these results, demonstrating the robustness of our findings (Fig.~\ref{fig:benchmarking_with_10x}).
In contrast to existing PLMs, SCEPTR performs on par with or better than TCRdist (Figs.~\ref{fig:benchmarking_overview}d / \ref{fig:individual_rocs}, Table~\ref{tab:per_epitope_aurocs_200_shot}).
For a reference set of size 200, SCEPTR performs better than TCRdist for five out of six tested peptides ($p=0.11$, binomial test), and for all six peptides when compared to all other models ($p=0.015$, binomial test).

We additionally compared models using the average distance between a query TCR and all references, instead of only the nearest neighbour (Fig.~\ref{fig:benchmarking_avg_dist}).
In this case, SCEPTR outperforms other models by an even wider margin.
Interestingly, all models perform worse compared to their nearest neighbour counterpart.
This finding might be explained mechanistically by the multiplicity of viable binding solutions with distinct sequence-level features, which are thought to make up pMHC-specific TCR repertoires~\cite{mayer_measures_2023,henderson_limits_2024}.
We hypothesized that the poor performance of prior PLMs might be overcome by learning projections from their high-dimensional representation space that align better with the TCR co-specificity prediction task.
We thus used the embeddings as input to linear support vector classifiers (Appendix~\ref{sec:transfer_learning_with_svc}).
Optimised linear probing does improve prediction performance (Fig.~\ref{fig:svc_benchmark_full}), but they remain inferior to nearest neighbour predictions using SCEPTR further illustrating the usefulness of SCEPTR embeddings for data-efficient transfer learning.

In some use cases, we may want to apply SCEPTR to the analysis of single chain TCR data. 
In benchmarking on either $\alpha$ or $\beta$ chain reference data alone prediction accuracy drops somewhat regardless of distance measure (Fig.~\ref{fig:ablation_chain}), as expected given that both receptor chains provide non-redundant information about specificity \cite{henderson_limits_2024}. Importantly though, SCEPTR distances also provide comparable or better prediction accuracy than TCRdist on the single chain level.

\begin{figure*}
  \includegraphics{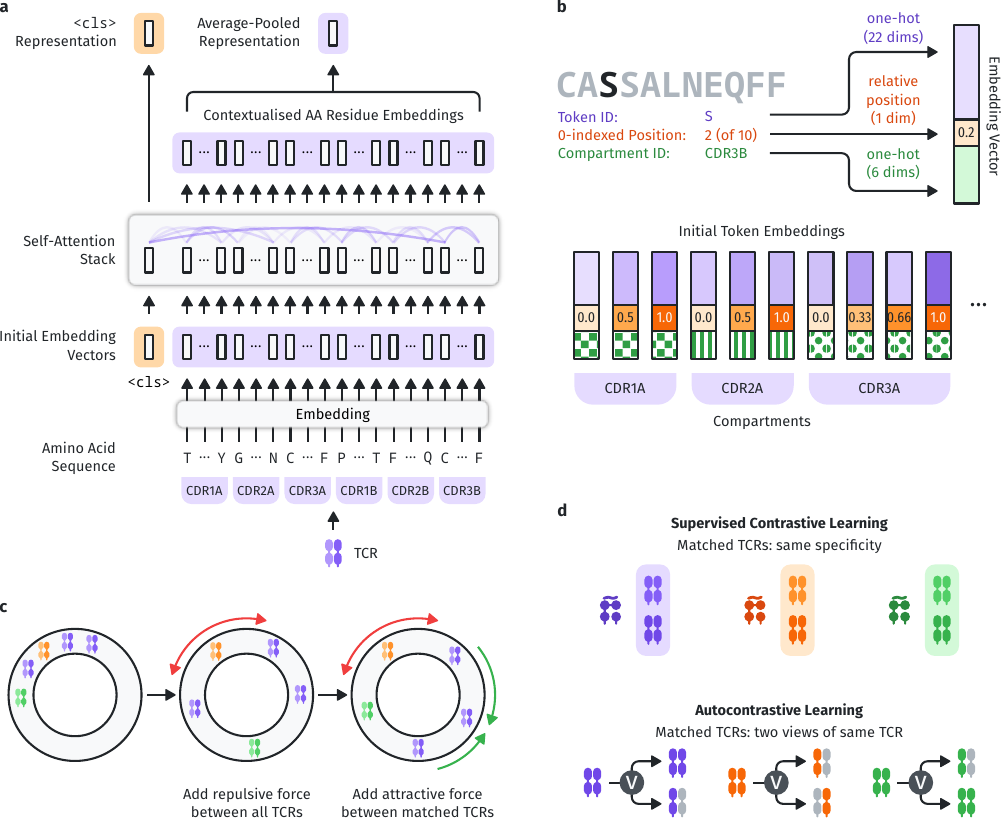}
  \caption{
    \textbf{A visual introduction to how SCEPTR works.}
    \textbf{a)}~SCEPTR featurises an input TCR as the amino acid sequences of its six CDR loops.
    Each amino acid residue is vectorised to $\reals^{64}$ (see panel \textbf{b}) and are passed along with the special \cls token vector through a stack of three self-attention layers.
    SCEPTR uses the contextualised embedding of the \cls token as the overall TCR representation, in contrast to the average-pooling representations used by other models.
    \textbf{b)}~SCEPTR's initial token embedding module uses a simple one-hot system to encode a token's amino acid identity and CDR loop number, and allocates one dimension to encode the token's relative position within its CDR loop as a single real-valued scalar.
    \textbf{c)}~Contrastive learning allows us to explicitly optimise SCEPTR's representation mapping for TCR co-specificity prediction.
    At a high level, contrastive learning encourages representation models to make full use of the available representation space while keeping representations of similar input samples close together.
    \textbf{d)}~Contrastive learning generalises to both the supervised and unsupervised settings.
    In the supervised setting, positive pairs can be generated by sampling pairs of TCRs that are known to bind the same pMHC.
    In the unsupervised setting, positive pairs can be generated by generating two independent ``views'' of the same TCR.
    We implement this by only showing a random subset of the input data features for every view -- namely, we remove a proportion of input tokens and sometimes drop the $\alpha$ or $\beta$ chain entirely (see methods~\ref{sec:methods_sceptr_pretraining}).
  }
  \label{fig:about_sceptr}
\end{figure*}

\subsection{Autocontrastive learning as a pre-training strategy}
\label{sec:acl_pretraining}

We now briefly summarize SCEPTR's architecture and autocontrastive pre-training strategy (see \hyperref[sec:methods]{Methods} for full details).
SCEPTR featurises an input TCR as the amino acid sequences of its six CDR loops.
It uses a simple one-hot encoding system to embed the amino acid tokens, and uses a stack of three self-attention layers to generate a 64-dimensional representation vector of the input receptor (Fig.~\ref{fig:about_sceptr}a,b).
Unlike existing TCR language models, SCEPTR is jointly pre-trained using autocontrastive and masked-language modelling (MLM) (Fig.~\ref{fig:about_sceptr}c,d).

To motivate the considerations that have led us to adopt this training paradigm, some background on transformer architectures and their training by MLM needs to be introduced. 
The transformer is a neural network developed in NLP that uses dot-product attention to flexibly learn long-range dependencies in sequential data~\cite{vaswani_attention_2017}.
BERT is an encoder-only variation of the transformer useful for text analysis and processing~\cite{devlin_bert_2019}.
BERT's innovation was its ability to be pre-trained in an unsupervised manner through MLM, where snippets of text are fed to the model with a certain proportion of tokens (e.g. words) masked, and the model must use the surrounding context to reconstruct the masked tokens.
MLM allowed BERT and its derivative models to exploit large volumes of unlabelled data to learn grammar and syntax and achieve high performance on downstream textual tasks with comparatively little supervised fine-tuning.

While MLM-trained PLMs have been successful in some protein prediction tasks \cite{rives2021biological,lin_evolutionary-scale_2023,elnaggar_prottrans_2022}, they have been documented to struggle with others \cite{li_feature_2024}.
Our benchmarking results led us to believe that MLM pre-training may not be optimal for TCR-pMHC specificity prediction.
Firstly, the majority of observed TCR sequence variation is attributable to the stochastic process of VDJ recombination.
As such, MLM may not teach models much transferable knowledge for specificity prediction.
Secondly, since the low volume of specificity-labelled TCR data provides limited opportunities for fine-tuning complex models, representation distances should ideally be directly predictive of co-specificity.

We were inspired to use contrastive learning to overcome these problems by the success of our previous work using statistical approaches to uncover patterns of sequence similarity characteristic of ligand-specific TCR repertoires \cite{mayer_measures_2023,tiffeau2024unbiased,henderson_limits_2024}.
Contrastive learning minimises distances between model representations of positive sample pairs while maximising distances between background pairs (Fig.~\ref{fig:about_sceptr}c) through a loss function of the following form~\cite{wang_understanding_2022,khosla_supervised_2021}:
\begin{multline}
  \Loss{contrastive}(f) \coloneqq \\
  \E_{
    \substack{
      (x, x^+) \sim \p{pos} \\
      \{y_i\}_{i=1}^{N} \overset{\mathrm{iid}}{\sim} \p{data}
    }
  }
  \left[
    -\log\frac
    {e^{f(x)^\top f(x^+)}}
    {
      e^{f(x)^\top f(x^+)} +
      \sum_{i}{e^{f(x)^\top f(y_i)}}
    }
    \right]
  \label{eq:contrastive}
\end{multline}
where $f: \mathcal{X} \to \uhsphere^{m-1}$ is a trainable embedding mapping from sample observation space $\mathcal{X}$ to points on the m-dimensional unit hypersphere $\uhsphere^{m-1} \subset \reals^m$, $\p{pos}$ is the joint distribution of positive pairs, $\p{data}$ is the overall data distribution, and $N \in \integers_+$ is some fixed number of background samples.

There are several well-known variants of this learning approach. In supervised contrastive learning, positive pairs are generated by sampling observations known to belong to the same class (Fig.~\ref{fig:about_sceptr}d top).
In the context of TCRs, we can define positive pairs to be TCRs annotated to interact with the same pMHC, in which case contrastive learning regresses distances between TCR pairs to their probabilities of co-specificity.
Autocontrastive learning approximates such positive pairs through data augmentation by generating two independent ``views''' of the same observation (Fig.~\ref{fig:about_sceptr}d bottom).

Given the scarcity of available labelled data, we opted to use the autocontrastive approach for purely unsupervised PLM pre-training (see Sec.~\ref{sec:scl_finetuning} for an  application of supervised contrastive learning, more similar to other recent applications of contrastive learning to TCRs \cite{drost_metcrs_2022,pertseva_tcr_2024}).
For this pre-training, we used data on close to a million unique paired TCRs obtained by Tanno et al. \cite{tanno_determinants_2020}, which represents one the largest collections of $\alpha\beta$ TCRs from a single study collected to date.
This data was filtered and standardized as described in methods~\ref{sec:methods_sceptr_pretraining}.
We generate different ``views'' of a TCR by dropout noise as is standard in NLP \cite{gao_simcse_2022}, but additionally adopted a censoring strategy inspired by masked-language modeling that randomly removes a proportion of residues or even complete $\alpha$ or $\beta$ chains.
In contrast to the only other study known to us having explored the application of autocontrastive learning to TCRs \cite{fang_attention-aware_2022}, we trained SCEPTR on all six hypervariable loops of the full paired chain $\alpha\beta$ TCR, as all contribute to TCR-pMHC specificity \cite{henderson_limits_2024}.
That being said, our chain dropping procedure during censoring ensures that single chain data are also in distribution for the model, giving SCEPTR flexibility for downstream applications with bulk sequenced TCR repertoires (Fig.~\ref{fig:ablation_chain}).

We define SCEPTR's output representation vector to be a contextualised embedding of a special input token called \cls (the naming convention for \cls comes from the fact that the output of this vector is often used for downstream \emph{classification}~\cite{devlin_bert_2019}), which is always appended to the tokenised representation of an input TCR (Fig.~\ref{fig:about_sceptr}a).
This allows SCEPTR to fully exploit the attention mechanism when generating the overall TCR representation.
Such training of a sequence-level representation is uniquely made possible by having an objective -- the contrastive loss (Eq.~\ref{eq:contrastive}) -- that directly acts on the representation output.
In contrast, MLM-trained PLMs such as ProtBert, ESM2 and TCR-BERT generate sequence embeddings by average-pooling the contextualised embeddings of each input token at some layer: a destructive operation which risks diluting information~\cite{li_feature_2024}.

\begin{figure*}
  \includegraphics{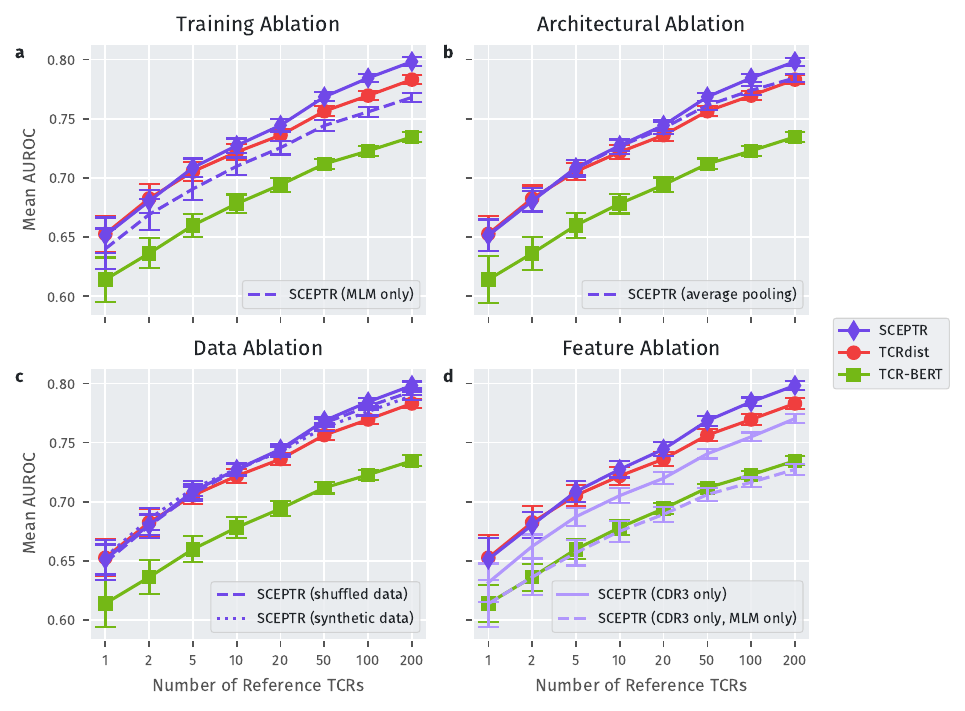}
  \caption{
    \textbf{Autocontrastive pre-training significantly improves SCEPTR's downstream performance.}
    The subplots show performance profiles of SCEPTR, TCRdist, TCR-BERT, and various ablation variants of SCEPTR on binary specificity prediction.
    \textbf{a)}~Training SCEPTR solely on MLM results in worse specificity prediction performance.
    \textbf{b)}~The baseline SCEPTR variant which uses the \cls pooling method performs marginally better than the variant which uses the average-pooling method. However, the average-pooling variant still performs on par with TCRdist.
    \textbf{c)}~Replacing SCEPTR's pre-training dataset with 1) the same dataset from Tanno et al., but with $\alpha/\beta$ chain pairing shuffled, and 2) synthetic data generated by OLGA both result in similar specificity prediction performance.
    \textbf{d)}~Restricting SCEPTR's featurisation of input TCRs to the amino acids of the $\alpha$ and $\beta$ CDR3 loops significantly worsen downstream performance.
    Additionally restricting training to only MLM further degrades performance, and produces a model with a near-equivalent performance profile to TCR-BERT.
  }
  \label{fig:ablation}
\end{figure*}

\subsection{Ablation studies}
\label{sec:ablation}

To understand which modelling choices drive the improved performance of SCEPTR, we trained variants of SCEPTR ablating a single component of either its architecture or training at a time, and benchmarked them using the framework described previously.

To establish the contribution of the autocontrastive learning to SCEPTR's performance, we trained the same model only using masked-language modelling:
\begin{description}
  \item[SCEPTR (MLM only)]
    This variant is trained only on MLM, without jointly optimising for autocontrastive learning.
    Following convention in the transformer field~\cite{li_sentence_2020}, TCR representation vectors are generated by average-pooling the contextualised vector embeddings of all constituent amino acid tokens produced by the penultimate self-attention layer, and $\ell$2-normalising the result.
\end{description}

The MLM-only variant underperforms compared to both SCEPTR and TCRdist, demonstrating that autocontrastive learning is a necessary ingredient for the increased performance of SCEPTR in few-shot specificity prediction (Fig.~\ref{fig:ablation}a).

We next sought to determine how much our pooling strategy and training dataset choice contributed to SCEPTR's performance gain.
First, we asked whether autocontrastive learning also improves embeddings generated via token average-pooling:

\begin{description}
  \item[SCEPTR (average pooling)]
    This variant receives both autocontrastive learning and MLM, but uses the average-pooling method to generate TCR representations.
\end{description}

While SCEPTR's \cls embeddings achieve the best results, the autocontrastive average-pooling variant still performs on par with TCRdist (Fig.~\ref{fig:ablation}b).

Second, we determined how the performance of SCEPTR depends on the precise dataset used for pre-training. To answer this question we trained two variants of SCEPTR using size-matched datasets:
\begin{description}
  \item[SCEPTR (synthetic data)]
    This variant is trained on a size-matched set unlabelled $\alpha\beta$TCRs generated by OLGA~\cite{sethna2019olga}, a probabilistic model of VDJ recombination.
    This synthetic data models only the recombination statistics and thus estimates the TCR distribution without taking into account the imprints of thymic and peripheral selection found in real repertoires \cite{mayer_measures_2023}.
  \item[SCEPTR (shuffled data)]
    This variant is trained on the same set of $\alpha\beta$TCRs as the original model, but the $\alpha$/$\beta$ chain pairing is randomised, thus removing pairing biases \cite{milighetti2024intra}.
\end{description}

We find that SCEPTR trained on synthetic or shuffled data performs worse for five out of six pMHCs ($p=0.11$, binomial test), but differences in AUROCs between model variants are small and regardless of training data SCEPTR performs on par with TCRdist (Fig.~\ref{fig:ablation}c).

Taken together, these ablation studies provide evidence that autocontrastive learning is the main factor enabling SCEPTR to close the gap between PLMs and alignment-based methods.

Information-theoretic analysis of the sequence determinants of TCR specificity demonstrate that all CDR loops and their pairing are important for determining binding specificity~\cite{henderson_limits_2024}.
To understand how much SCEPTR's improved performance with respect to TCR-BERT is due to the restriction of the latter model's input to the CDR3 alone, we trained variants of SCEPTR restricted to this hypervariable loop:

\begin{description}
  \item[SCEPTR (CDR3 only)]
    This variant only accepts the $\alpha$ and $\beta$ chain CDR3 sequences as input (without knowledge of the V genes/first two CDR loops of each chain).
    It is jointly optimised for MLM and autocontrastive learning.
  \item[SCEPTR (CDR3 only, MLM only)]
    This otherwise equivalent variant is only trained using the MLM objective, and thus uses the average-pooling representation method.
\end{description}

The results demonstrate that taking into account all CDR loops leads to a performance gain as expected (Fig.~\ref{fig:ablation}d).
We also see that autocontrastive learning even when restricted to CDR3s leads to a substantial performance gain, helping the autocontrastive CDR3 variant achieve similar performance to the full-input MLM-only variant.

\begin{figure*}
  \includegraphics{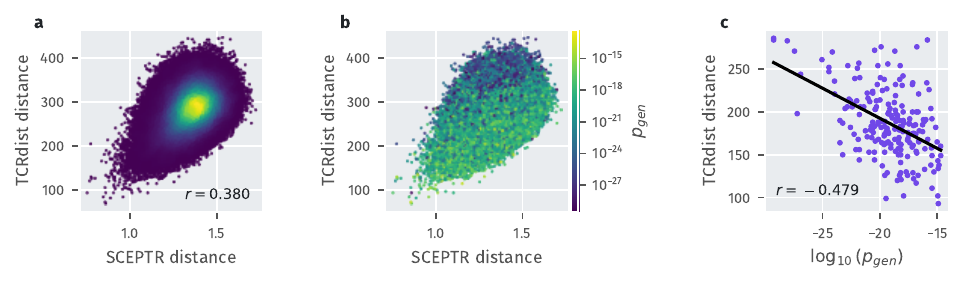}
  \caption{
    \textbf{SCEPTR embedding distances weight sequence similarity with respect to recombination biases.}
    \textbf{a)}~Scatter plot of SCEPTR and TCRdist distances between pairs of TCRs from the held-out test set of the pre-training dataset. The points are coloured according to a Gaussian kernel density estimate.
    \textbf{b)}~Colouring TCR pairs instead by the minimal probability of generation $\p{gen}$ of the two TCRs as estimated by OLGA~\cite{sethna2019olga} suggests that SCEPTR embeddings locally contract regions of representation space that due to recombination biases are sparsely sampled.
    \textbf{c)}~For sequence pairs judged to be similar by SCEPTR (distance $\in [0.98, 1.02]$), variations in $\p{gen}$ explain a substantial fraction of the variance in TCRdist, providing statistical evidence for the hypothesized weighting of sequence similarity with respect to the local density of sequences produced by VDJ recombination (see Fig.~\ref{fig:pgen_vs_tcrdist} for the generality of this dependence across SCEPTR bins).
  }
  \label{fig:sceptr_vs_tcrdist}
\end{figure*}

\subsection{Comparison of SCEPTR embeddings to alignment-based TCR similarity}

To gain insights into what SCEPTR has learned during pre-training, we compared its embedding distances to alignment-based distances as calculated by TCRdist.
To do so, we calculated all pairwise distances between one thousand TCRs randomly sampled from the testing partition of the Tanno et al. dataset~\cite{tanno_determinants_2020} (held out during SCEPTR pre-training) using both models.
We find that SCEPTR and TCRdist distances are clearly correlated (Fig.~\ref{fig:sceptr_vs_tcrdist}a).
This shows that in parts the success of SCEPTR can be understood by its embedding distances providing good alignment-free approximations to traditional sequence similarity measures.
Yet, there is also substantial variability between both measures for many pairs, and an inspection of such discordant pairs can provide insights into how metrics differ.

First, we noticed that pairs of sequences judged to be similar by SCEPTR but not TCRdist were less likely to be generated during VDJ recombination (Fig.~\ref{fig:sceptr_vs_tcrdist}b).
We found that among sequence pairs judged to be similar by SCEPTR, TCRdist distance was strongly negatively correlated with $\p{gen}$ (Figs.~\ref{fig:sceptr_vs_tcrdist}c / \ref{fig:pgen_vs_tcrdist}).
This implies that SCEPTR embeds TCRs closer to each other if they are in regions of sequence space that are less densely sampled by the generative distribution. 
As argued in detail in the \hyperref[sec:discussion]{Discussion}, this property of SCEPTR embeddings is expected on theoretical grounds due to the loss function used for contrastive learning. 
This property might enable SCEPTR embeddings to capture the intuition that finding close-by nearest neighbours is more surprising for sequences with low $\p{gen}$ and thus more informative compared to similarity between highly probable TCRs.

Second, we noticed that a high similarity on a single chain tended to be sufficient for a small SCEPTR distance (Fig.~\ref{fig:sceptr_vs_tcrdist_chain_delta}).
To quantify this effect, we analysed how SCEPTR distances correlate with different ways of averaging the $\alpha$ and $\beta$ chain TCRdist distances into a paired chain measure. We found that SCEPTR distances correlate more closely with the minimum distance of the two chains rather than their arithmetic mean  (Fig.~\ref{fig:sceptr_and_tcrdist_averaging.pdf}).
Further investigation might focus on whether this property helps prediction performance due to the varying contributions of the TCR $\alpha$ and $\beta$ chain to specific binding across pMHCs \cite{henderson_limits_2024}.

\begin{figure}
  \centering
  \includegraphics{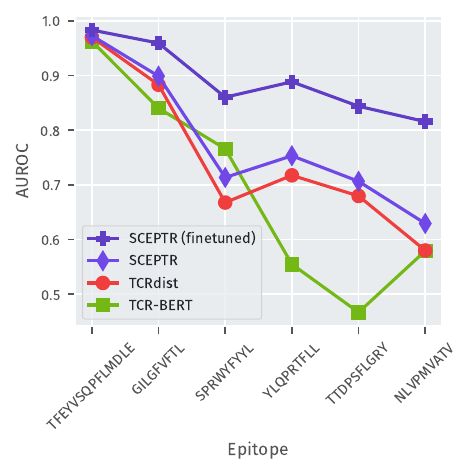}
  \caption{
    \textbf{Supervised contrastive learning improves discrimination between pMHCs.}
    Prediction performance as measured by AUROC on binary one-versus-rest classification for each of six pMHCs for different models. 
    The fine-tuned model improves performance by exploiting the discriminative nature of the classification task.
  }
  \label{fig:scl_benchmark}
\end{figure}

\subsection{Supervised contrastive learning as a fine-tuning strategy}
\label{sec:scl_finetuning}

Supervised contrastive learning provides an avenue to further optimise pre-trained embeddings for TCR specificity prediction. As a proof-of-concept, we fine-tuned SCEPTR to better discriminate between the six pMHC specificities used as the benchmarking targets in section~\ref{sec:benchmarking}.

For this task we took all the TCRs annotated against the target pMHCs from our labelled TCR dataset, and split them into a training, a validation, and a testing set.
We ensured that no study used for training or validation contributed any data to the test set, so that the fine-tuned model would not be able to achieve good performance simply by exploiting inter-dataset biases.
The training set included 200 binders against each target pMHC, totalling to 1200 TCRs.
The rest of the TCRs from the same studies were used to construct the validation set.
TCRs from all remaining studies were used for the testing set, which comprised of 5670 TCRs.
SCEPTR was fine-tuned on the training set with supervised contrastive learning, using the validation loss for early stopping (methods~\ref{sec:methods_sceptr_finetuning}).

We used the framework from section~\ref{sec:benchmarking} to benchmark the performance of fine-tuned SCEPTR, using the training set as the references.
The results show that fine-tuning can greatly improve the ability of the model to discriminate between pMHCs (Fig.~\ref{fig:scl_benchmark}).
Improvements are most noticeable for the pMHCs against which other methods achieve relatively low performance.
When filtering all TCRs with greater than 90\% or 80\% sequence similarity to any training sequence from the test set, the fine-tuned model still improves performance significantly  (Fig.~\ref{fig:scl_discrimination_filtering}) showing that learning goes beyond memorization of public TCRs.

\begin{figure}
  \includegraphics{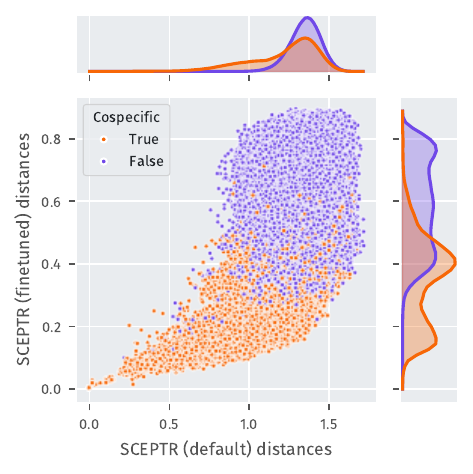}
  \caption{
    \textbf{Supervised contrastive learning reshapes the embedding space.}
    Scatter plot of distance between pairs of TCRs as measured by the pre-trained and fine-tuned versions of SCEPTR, and their marginal conditional probability density functions.
    Points are coloured based on whether the corresponding TCR pair involves TCRs that are annotated against the same pMHC (orange) or against different pMHCs (purple).
    Pairs of TCRs were obtained by sampling one receptor from the training partition and the other from the test partition of the labelled TCR data used during fine-tuning.
  }
  \label{fig:pretrained_vs_finetuned_sceptr}
\end{figure}

Interestingly, unlike other models, fine-tuned SCEPTR makes better inferences by measuring the average distance between a query TCR and all reference TCRs instead only the nearest TCR (Fig.~\ref{fig:sceptr_baseline_finetuned_nn_vs_avg_dist}). This suggests an ability of supervised contrastive fine-tuning to help the model discover the commonalities between the multiple different binding solutions thought to exist for each pMHC.
We thus analysed how fine-tuning changes the SCEPTR embedding distances between co-specific and cross-pMHC TCR pairs (Fig.~\ref{fig:pretrained_vs_finetuned_sceptr}).
Unexpectedly, we found that a major difference of fine-tuned SCEPTR distances concerns cross-pMHC pairs.
We observe that fine-tuning allows the model to identify a subset of ``easy'' negative pairs. These presumably involve TCRs the model is highly confident are specific to different pMHC, thus illustrating how discrimination between a fixed set of potential target pMHCs is easier than binary classification with respect to TCRs of arbitrary specificity.
Conversely, the fine-tuned model's performance degrades with respect to unseen pMHCs (Fig.~\ref{fig:scl_unseen}), perhaps unsurprisingly given the very limited number of pMHCs represented in the training data.

Another notable feature of Fig.~\ref{fig:scl_benchmark} is that performance varies substantially by ligand.
That is, prediction is easier for some pMHCs, regardless of method.
This is a phenomenon that has also been observed in public benchmarks \cite{nielsen_lessons_2024}.
Consequently, we used coincidence (order two R\'enyi) entropy measures \cite{mayer_measures_2023,henderson_limits_2024} to discover intrinsic properties of the pMHC-specific TCR repertoires that determine model AUROCs.
We find that TCRs annotated against the ``easier'' pMHCs have lower V/J gene diversity (Fig.~\ref{fig:entropy_vs_performance}) and lower average sequence distances between pairs of TCRs (Fig.~\ref{fig:pc_vs_performance}).
This indicates that differences in models' ability to predict TCR-pMHC specificity are linked to the diversity of epitope-specific repertoires.

\section{Discussion}
\label{sec:discussion} 

In this study, we have introduced SCEPTR, a pre-trained TCR PLM that achieves state-of-the-art few-shot TCR-pMHC specificity prediction accuracy.
Through SCEPTR we demonstrate that joint autocontrastive and masked-language pre-training is a paradigm for learning PLMs better aligned with TCR specificity prediction tasks.
Our model can be readily used for alignment-free TCR analysis in downstream applications (see \hyperref[sec:code_availability]{code availability}) including the unsupervised discovery of antigen-specific T cell groups (metaclonotypes) by sequence-based clustering  \cite{dash_quantifiable_2017,mayer2021tcr}. 

A limitation of our study is that we did not undertake a complete exploration of training and architectural hyperparameters, as well as training dataset choice.
We envisage multiple avenues that may improve SCEPTR further.
Firstly, training could be made more efficient by optimising the distribution of masked/dropped tokens during pre-training, taking into account the variable relevance of different parts of the sequence in determining specificity~\cite{henderson_limits_2024}.
Secondly, as certain sequence motifs appear recurrently (e.g. CDR3 loops often begin with \texttt{CAS}), a more intelligent tokenisation scheme could offload learning of these primary sequence statistics into the tokenisation process.
Finally, with the emergence of increasingly large paired-chain TCR datasets \cite{spindler2020massively,raybould2024observed,sureshchandra2024tissue}, retraining SCEPTR on data from multiple sources could eliminate biases inherent to specific experimental approaches and donor MHC restrictions.

Pre-trained PLMs have achieved high performance on protein stability and structural predictions~\cite{lin_evolutionary-scale_2023, elnaggar_prottrans_2022}.
However, we find that existing PLMs fail to confer similar benefits to predicting TCR-pMHC interactions.
This finding adds to recent work showing that current PLM pre-training is not well-aligned with certain downstream tasks~\cite{li_feature_2024}.
Importantly, we show that autocontrastive pre-training can overcome misalignment, and thus provide a constructive path out of this impasse which could also be applied outside of the TCR domain.

What determines whether a certain downstream task is aligned with MLM pre-training?
MLM teaches PLMs to predict the conditional distribution of tokens given sequence context.
Thus it stands to reason that amenable downstream tasks involve predictions of properties that determine the distribution of observed proteins on sequence space.
Observed proteins tend to concentrate in areas of sequence space with higher protein stability since evolution on average selects for this property \cite{bloom2006protein}.
For datasets containing protein families whose members have a conserved structure despite primary sequence variation, co-evolutionary couplings driven by structural constraints influence allowed sequence variability \cite{weigt2009identification,jumper2021highly,lin_evolutionary-scale_2023}.
These data distributional properties might explain how MLM can teach PLMs features related to both stability and structure.

In contrast, the distribution of TCRs over sequence space is primarily shaped by the biases of VDJ recombination with antigen-specific selection playing an important, but likely second-order effect \cite{mayer_measures_2023}.
While long-term evolutionary pressures may act to align recombination statistics with TCR function \cite{Mayer2015HowWellAdapted,thomas2019selected}, empirical evidence so far suggests recombination biases primarily anticipate thymic selection for stability and folding \cite{Elhanati2014QuantifyingSelection}.
In contrast, studies to date have found no clear relationship between probabilities of recombination and the likelihood of receptors engaging specific pMHCs \cite{sethna2019olga}.
Given these considerations, we expect MLM pre-training to align better to tasks concerning VDJ recombination than to TCR specificity prediction.
Indeed, previous work training PLMs on adaptive immune receptors has demonstrated that embeddings strongly depend on V/J gene usage and can be used to predict primarily generation-related properties such as receptor publicity \cite{GoldnerKabeli2024SelfsupervisedLearning,wang2024language}.

Why does autocontrastive learning help to generate embeddings better suited for specificity prediction?
An interesting insight comes from an asymptotic decomposition of the contrastive loss function into the \emph{uniformity} and \emph{alignment} terms \cite{wang_understanding_2022}:

\begin{align}
  \textrm{Unif.}(f) &\coloneqq
  \log{
    \mathop{\mathbb{E}}_{x,y \overset{\mathrm{iid}}{\sim}
    \p{data}}\left[
      e^{-\lVert f(x) - f(y) \rVert^2}
      \right]
  } \label{eq:uniformity} \\
  \textrm{Align.}(f) &\coloneqq
  \mathop{\mathbb{E}}_{(x, x^+) \sim \p{pos}}\left[
    \lVert f(x) - f(x^+) \rVert
    \right] \label{eq:alignment}
\end{align}

Uniformity incentivises the model to make use of the full representation space, while alignment minimises the expected distance between positive pairs (e.g. co-specific TCRs)~\cite{wang_understanding_2022}.
From this view, contrastive learning on adaptive immune receptor data encourages PLMs to undo the large-scale distributional biases created by VDJ recombination through the uniformity term, while helping to identify features relating to TCR (co-)specificity via the alignment term.
While \emph{auto}contrastive learning approximates the alignment term through the generation of pairs of views, it still provides a direct empirical estimate for the uniformity term.
Thus, a key benefit of autocontrastive learning may be that it reduces the confounding effects of VDJ recombination in embedding space.
SCEPTR's ability to ``adjust'' its distances for $\p{gen}$ as demonstrated in Figs.~\ref{fig:sceptr_vs_tcrdist} and \ref{fig:pgen_vs_tcrdist} lends support to this conjecture.

\begin{figure}
  \includegraphics{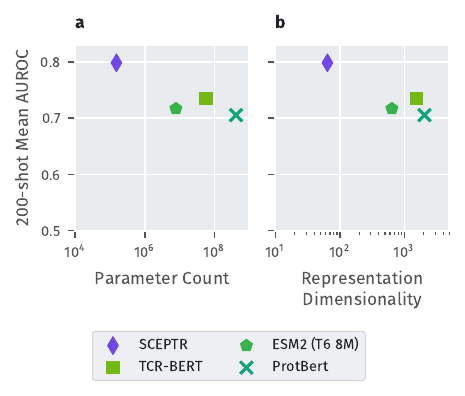}
  \caption{
    \textbf{Model complexity does not correlate with downstream performance.}
    Model performance as measured by mean 200-shot AUROC (section~\ref{sec:benchmarking}) does not scale with model complexity as measured by either \textbf{a)} parameter count or \textbf{b)} representation dimensionality.
    Despite being the smallest PLM by a wide margin, SCEPTR performs better than alternative models.
  }
  \label{fig:model_complexity_vs_performance}
\end{figure}

Comparing SCEPTR to other PLMs suggests that model complexity as measured by either parameter count or representation dimensionality is not currently the limiting factor for TCR-pMHC prediction performance (Fig.~\ref{fig:model_complexity_vs_performance}).
This is directly supported by how the CDR3-only, MLM-only variant of SCEPTR performs almost equivalently to the much larger but similarly pre-trained TCR-BERT (Fig.~\ref{fig:ablation}d).
This finding stands in contrast to observations of performance scaling with model size in general PLMs~\cite{lin_evolutionary-scale_2023} and antibody language modelling~\cite{barton_generative_2024}.
While the focus of the current study was on training a simple model, it would be interesting in future work to investigate performance scaling with model complexity and training dataset size with our novel training procedure.

Looking forward, there are many exciting avenues to further develop contrastive learning as a paradigm to crack the TCR code.
For example, there may be ways to exploit the uniformity (Eq.~\ref{eq:uniformity}) and alignment (Eq.~\ref{eq:alignment}) decomposition to simultaneously train on unlabelled and specificity-labelled data. A practical benefit of our contrastive learning formulation is that it does not require any optimisation with respect to the true negative distribution (i.e. TCRs that are explicitly not co-specific) -- a non-trivial distribution to estimate for TCRs~\cite{dens_pitfalls_2023}.

Another interesting avenue is the use of labels other than pMHC specificity -- such as phenotypic annotations from single-cell data -- as additional supervised contrastive training signals.
While supervised contrastive learning does not currently lead to generalisable learning beyond training pMHCs, we expect a transition towards generalisation as larger volumes of specificity-labelled TCR data become available, as has been the case with supervised contrastive learning in other fields \cite{schroff_facenet_2015, liu_sphereface_2017, chen_mobilefacenets_2018, deng_arcface_2021, gao_simcse_2022}.

Finally, while the focus of this work given current data limitations has been on learning TCR embeddings, contrastive learning may also help us learn effective joint TCR-pMHC embeddings in the future when the joint space (and particularly the pMHC space) is better sampled, and thus ultimately enable the zero-shot prediction of TCR-pMHC specificity.

\section{Methods}
\label{sec:methods}

\subsection{Model benchmarking}
\label{sec:methods_benchmarking}

For each pMHC, we varied the number $k$ of reference TCRs where $k \in \{1, 2, 5, 10, 20, 50, 100, 200\}$.
Within each model-pMHC-$k$-shot combination, we benchmarked multiple reference-test splits of the data to ensure robustness.
For $k=1$, we benchmarked every possible split.
For $k \in [2,200]$, we benchmarked 100 random splits, where we ensured that the same splits were used across all models to reduce extraneous variance.

In assessing the statistical significance of differences in average model performance, we took a paired difference approach. 
We expected certain pMHCs and data splits to present a more difficult prediction problem than others.
As we are interested in assessing the relative performance of models, we calculated the variance across splits of the difference between each individual model's AUROC and the average across all models.
For each model, we estimated this variance within each of the pMHCs, and then averaged these variances to obtain an estimate of overall variance.

The CDR3 Levenshtein model computes the distance between two TCRs as the sum of the Levenshtein distances between the receptors' $\alpha$ and $\beta$ CDR3s.

Note that while SCEPTR's architecture and training allows it to directly generate representation vectors for complete $\alpha\beta$ TCR sequences (Fig.~\ref{fig:about_sceptr}a, methods~\ref{sec:methods_sceptr_architecture}), this is not the case for the other PLMs.
For TCR-BERT, ESM2 and ProtBert representations for the $\alpha$ and $\beta$ chains were independently generated, then concatenated together, and finally average-pooled to produce an embedding of the heterodimeric receptor (see appendix~\ref{sec:using_existing_protein_language_models}).

\subsection{SCEPTR architecture}
\label{sec:methods_sceptr_architecture}

\textbf{SCEPTR} (\SCEPTRFull) is a BERT-like transformer encoder that maps TCR sequences to vector embeddings.
Like BERT, it is comprised of a tokeniser module, an embedder module and a self-attention stack (Fig.~\ref{fig:about_sceptr}a).

The tokeniser module represents each input TCR as the amino acid sequences of the first, second and third complementarity-determining regions (CDRs) of each chain, where each amino acid is a token.
A special \cls token is appended to each input TCR, as its contextualised embedding will eventually become SCEPTR's output representation vector (Fig.~\ref{fig:about_sceptr}a,b).

SCEPTR uses a simple, non-trainable embedder module, where a one-hot vector is used to encode token identity (22 dimensions for 20 amino acids plus special tokens \cls and \mask), and token positions are specified by first one-hot encoding the containing CDR loop number (6 dimensions), then encoding the token's relative position within the loop as a single scalar variable (Fig.~\ref{fig:about_sceptr}b).
This results in initial token embeddings in $\reals^{29}$, which are passed through a trainable linear projection onto $\reals^{64}$. SCEPTR's self-attention stack then operates at this fixed dimensionality (Fig.~\ref{fig:self_attention_stack}).
SCEPTR's self-attention stack comprises three layers, each with eight attention heads and a feed-forward dimensionality of 256, and is thus substantially simpler than existing models.
Our tests suggest that relative position embedding helps SCEPTR learn better calibrated TCR co-specificity rules (see appendix~\ref{sec:relative_position_embedding}).

\subsection{SCEPTR Pre-training}
\label{sec:methods_sceptr_pretraining}

\subsubsection{Data}

The unlabelled paired-chain $\alpha\beta$TCR sequences used to pre-train SCEPTR were taken from a study by Tanno et al.~\cite{tanno_determinants_2020}, which provides 965,523 unique clonotypes sampled from the blood of 15 healthy human subjects.
As opposed to traditional single-cell sequencing, Tanno et al. used a ligation-based sequencing method to resolve which $\alpha$ chains paired with which $\beta$ chains.
To mitigate potential noise from incorrect chain pairing, we applied an extra processing step to remove clonotypes that shared the same nucleotide sequence for either the $\alpha$ or the $\beta$ chain, as previously described \cite{mayer_measures_2023}.
After filtering for functional TCRs using tidytcells, a TCR gene symbol standardiser~\cite{nagano_tidytcells_2023}, we retained 842,683 distinct clonotypes.

A random sub-sample of 10\% of this data was reserved for use as an unseen test set, containing 84,268 unique clonotypes distributed across 83,979 unique TCRs.
Of the remaining 90\% of the data, we filtered out any clonotypes with amino acid sequences that also appeared in the test set, resulting in a training set of 753,838 unique clonotypes across 733,070 unique TCRs.

\subsubsection{Procedure}

SCEPTR was jointly optimised for MLM and autocontrastive learning, where the total loss of a training step was calculated as the sum of the MLM and autocontrastive (Eq.~\ref{eq:autocontrastive_loss}) losses.

We implemented MLM following established procedures~\cite{devlin_bert_2019}.
Namely, 15\% of input tokens were masked, and masked tokens had an 80\% probability of being replaced with the \mask token, a 10\% probability of being replaced by a randomly chosen amino acid distinct from the original, and a 10\% probability of remaining unchanged.
The MLM loss was computed as the cross-entropy between SCEPTR's predicted token probability distribution and the ground truth.

Our choice of autocontrastive loss function is inspired by related work in NLP~\cite{gao_simcse_2022} and computer vision~\cite{khosla_supervised_2021}, but adapted to the TCR setting.
Let $B = \{\tcr_i\}_{i=1}^N$ be a minibatch of $N$ TCRs.
We generate two independent ``views'' of each TCR $\tcr_i$ by passing two censored-variants of the same receptor through the model.
Our censoring procedure removes a random subset of a fixed proportion (20\%) of the residues from the tokenised representation of the CDR loops and with a 50\% chance drops either the full $\alpha$ or $\beta$ chain.
To ensure that censoring does not fundamentally alter the underlying TCR sequence, the positional encoding for each token remains fixed relative to the original TCR.
In addition to the random censoring, views also differ due to dropout noise during independent model passes.
Taken together, this procedure maps the minibatch $B$ to the set of $2N$ TCR views $V = \{v_j\}_{j=1}^{2N}$, where $v_{2i}$ and $v_{2i-1}$ are two independent views of the same TCR $\tcr_i$ ($i \in \{1 ... N\}$).
Where $k \in I = \{1 ... 2N\}$ is an arbitrary index of a view $v_k \in V$, let $p(k)$ be the index of the other view generated from the same TCR, and $N(k) = \{l \in I: l \not = k\}$ be the set of all indices apart from $k$ itself.
Let $\mathbf{r}_k$ denote SCEPTR's vector representation of TCR view $v_k$.
Then the autocontrastive loss for minibatch $B$ is computed as follows:

\begin{equation}
  \Loss{AC}(B)
  = \frac{1}{2N} \sum_{k \in I}
  -\log \frac
  { e^{\mathbf{r}_k^\top \mathbf{r}_{p(k)} / \tau} }
  { \sum_{n \in N(k)} e^{ \mathbf{r}_k^\top \mathbf{r}_n / \tau}}
  \label{eq:autocontrastive_loss}
\end{equation}

Here, $\tau$ is a temperature hyper-parameter which we set to 0.05 during training, following previous literature~\cite{gao_simcse_2022}.

We used ADAM (adaptive moment estimation)~\cite{kingma_adam_2017} to perform stochastic gradient descent.
We chose a minibatch size of 1024 samples and trained for 200 epochs, which equated to 143,200 training steps.
The internal dropout noise of SCEPTR's self-attention stack was set to 0.1.

Our methodology of randomly censoring residues and even entire chains stands in contrast to previous work in NLP by \citet{gao_simcse_2022}, who found that relying only on the internal random drop-out noise of the language model was sufficient for effective autoconstrastive learning.
However, our experiments suggest that in the TCR domain, residue and chain censoring leads to embeddings with better downstream TCR specificity prediction performance (Fig.~\ref{fig:ablation_chain}).

\subsection{SCEPTR fine-tuning with supervised contrastive learning}
\label{sec:methods_sceptr_finetuning}

\subsubsection{Data}

For supervised contrastive fine-tuning we took all TCR binders against the six best-sampled pMHC targets from our labelled TCR dataset, and split them into a training, a validation, and a test set such that no study used to construct the training or validation sets contributed any TCRs to the test set (Table~\ref{tab:data_split_studies}).

\subsubsection{Procedure}

The fine-tuning process involved the joint optimisation of SCEPTR on MLM and supervised contrastive learning.
As during pre-training, the overall loss for each training step was computed as the unweighted sum of the MLM and supervised contrastive (Eq.~\ref{eq:supervised_contrastive_loss}) losses.
The pre-trained state of SCEPTR was used as the starting point for fine-tuning.
With only 200 TCRs for each target pMHC to train on, we limited the number of learnable parameters by only allowing the weights of the final self-attention layer to be trainable.
Additionally, we monitored increases in validation loss for early stopping of fine-tuning, which occurred after 2 epochs, where one epoch is defined as the model seeing 100,000 binders for each pMHC.
Given our a batch size of 1,024 TCRs, this corresponded to a total of 1,172 training steps.

Our implementation of supervised contrastive learning closely follows the formulation suggested by Khosla et al.~\cite{khosla_supervised_2021}.
This approach to supervised contrastive learning combines loss contributions from true positive pairs, with those from second views of each positive instance (as in autocontrastive learning) as well as all views of all other sample points with the same pMHC label.
Let $B = \{\tcr_i\}_{i=1}^N$ be a minibatch of $N$ pMHC-annotated TCRs.
We use the same procedure as in our autocontrastive framework (see methods~\ref{sec:methods_sceptr_pretraining}) to generate two views of each of the TCRs, producing a set of $2N$ views $V = \{v_j\}_{j=1}^{2N}$.
Let $Y = \{y_i\}_{i=1}^N$ be the index-matched pMHC labels for TCRs in $B$, and $\bar{y}_j$ denote the labels mapped to the indices of the views in $V$ such that $\bar{y}_{2i} = \bar{y}_{2i-1} = y_i$.
Now given arbitrary sample view index $k$, let $P(k) = \{l \in A(k): \bar{y}_l = \bar{y}_k\}$ be the set of all indices whose corresponding samples have the same pMHC label as $v_k$, with cardinality $|P(k)|$.
The supervised contrastive loss for TCR minibatch $B$ is:
\begin{multline}
  \Loss{SC}(B) = \\
  \frac{1}{2N} \sum_{k \in I}
  \frac{1}{|P(k)|} \sum_{p \in P(k)}
  -\log \frac
  { e^{\mathbf{r}_k^\top \mathbf{r}_p / \tau} }
  { \sum_{n \in N(k)} e^{ \mathbf{r}_k^\top \mathbf{r}_n / \tau}}
  \label{eq:supervised_contrastive_loss}
\end{multline}
Each batch during fine-tuning has an equally balanced number of binders to each of the six pMHCs.

\acknowledgments

The authors thank Ned Wingreen, Chris Watkins, Trevor Graham, Sergio Quezada, Machel Reid, Linda Li, Rudy Yuen, Sankalan Bhattacharyya, and Matthew Cowley for useful discussions. YN and MM were supported by Cancer Research UK studentships under grants BCCG1C8R and A29287, respectively. The work of ATM was supported in parts by funding by the Royal Free Charity. 

The authors declare no competing interests.

\onecolumngrid
\section*{Code Availability}
\label{sec:code_availability}

\begin{description}
  \item[\url{https://github.com/yutanagano/sceptr}] a readily usable deployment of SCEPTR and its variants.
  \item[\url{https://github.com/yutanagano/tcrlm}] houses code used for designing and training our models.
  \item[\url{https://github.com/yutanagano/libtcrlm}] library code powering the above repositories
\end{description}
\twocolumngrid

\bibliographystyle{apsrev4-2}
\bibliography{bibliography.bib}

\clearpage
\onecolumngrid
\appendix

\setcounter{figure}{0}
\setcounter{table}{0}
\renewcommand{\thefigure}{S\arabic{figure}}
\renewcommand{\theHfigure}{S\arabic{figure}}
\renewcommand{\thetable}{S\Roman{table}}
\renewcommand{\theHtable}{S\Roman{table}}

\section{Generating TCR vector embeddings using existing protein language models}
\label{sec:using_existing_protein_language_models}

\subsection{TCR-BERT}

The TCR-BERT model was downloaded through HuggingFace at \url{https://huggingface.co/wukevin/tcr-bert}.
Since TCR-BERT is trained to read one CDR3 sequence at a time, we generated TCR representations by generating two independent representations of the $\alpha$ and $\beta$ chain, and concatenating them together.
The TCR-BERT representation of a chain was generated by feeding the model its CDR3 sequence, then taking the average pool of the amino acid token embeddings in the 8th self-attention layer, as recommended by the study authors~\cite{wu_tcr-bert_2021}.

\subsection{ESM2}

The ESM2 (T6 8M) model was downloaded through HuggingFace at \url{https://huggingface.co/facebook/esm2_t6_8M_UR50D}.
ESM2 is trained on full protein sequences, but not protein multimers.
Therefore, we generated ESM2 representations for the $\alpha$ and $\beta$ chains separately, and concatenated them to produce the overall TCR representation.
To generate the representation of a TCR chain, we first used Stitchr~\cite{heather_stitchr_2022} to reconstruct the full amino acid sequence of a TCR from its CDR3 sequence and V/J gene.
Then, the resulting sequence of each full chain was fed to ESM2.
We took the average-pooled result of the amino acid token embeddings of the final layer to generate the overall sequence representation, as recommended~\cite{lin_evolutionary-scale_2023}.

\subsection{ProtBert}

The ProtBert model was downloaded through HuggingFace at \url{https://huggingface.co/Rostlab/prot_bert}.
Similarly to ESM2, ProtBert is trained on full protein sequences.
Therefore, we again used Stitchr to generate full TCR chain amino acid sequences, and fed them to ProtBert to generate independent $\alpha$ and $\beta$ chain representations.
We again as recommended average-pooled the amino acid token embeddings of the final layer~\cite{elnaggar_prottrans_2022}.

\section{Learning features within embedding spaces}
\label{sec:transfer_learning_with_svc}

The focus of the current work has been to use nearest neighbour prediction using PLM embeddings as the most direct test of data-efficient transfer learning that works with as little as a single reference sequence.
If slightly more data is available, another approach is to train supervised predictors atop PLM embeddings.
To test how much such training can improve prediction performance, we trained linear support vector classifiers (SVC) on the PLM embeddings provided by different models.
In each instance, we trained the classifier to distinguish reference TCRs from 1000 randomly sampled background TCRs.
We outline the methodology in more detail below.

\begin{figure}
  \includegraphics{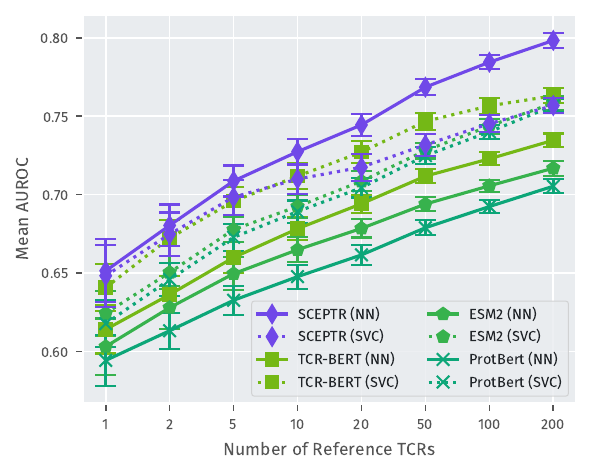}
  \caption{
    \textbf{Benchmarking linear support vector classifiers trained on PLM features on TCR specificity prediction.}
    Performance of different PLMs applied to few-shot TCR specificity prediction, either through nearest neighbour prediction (models marked as ``NN'' in the legend, see section~\ref{sec:benchmarking}) or using a linear support vector classifier trained atop their TCR featurisations (models marked as ``SVC'' in the legend, see appendix~\ref{sec:transfer_learning_with_svc}).
    For all PLMs except SCEPTR, training a linear SVC atop the model's features improves performance.
    SCEPTR (NN) outperforms all methods, even the SVC trained atop its own features.
  }
  \label{fig:svc_benchmark_full}
\end{figure}

We find that the SVC predictors for ProtBert, ESM2 and TCR-BERT all perform better than their nearest neighbour counterparts, but still worse than SCEPTR's nearest neighbour predictions (Fig.~\ref{fig:svc_benchmark_full}).
We also trained an SVC atop SCEPTR, which did not lead to further improvement upon the nearest neighbour prediction (Fig.~\ref{fig:svc_benchmark_full}).
These findings highlight how in the low data regime typical of most pMHCs, misalignment of pre-training to downstream tasks can only be partially remediated by training on reference TCRs. 

To train the linear SVCs on top of PLM features, we sampled 1000 random background TCRs from the training partition of the unlabelled Tanno et al. dataset.
We employed a similar strategy to our benchmarking in section~\ref{sec:benchmarking} to split our dataset of curated specificity-annotated $\alpha\beta$TCRs into a reference set and testing set.
For each PLM-pMHC-split combination, we trained a linear SVC using the PLM embeddings of the reference TCRs as the positives and those of the 1000 background TCRs as the negatives.
The same 1000 background TCRs were used across model-pMHC-split combinations to ensure consistency.
We accounted for the imbalance between the number of positive and negative samples used during SVC fitting by weighting the penalty contributions accordingly.
Finally, we tested SVCs using the same benchmarking classification task as previously described.

\section{Effects of different position embedding methods}
\label{sec:relative_position_embedding}

To better understand TCR similarity rules as learned by PLMs, we measured the average distance penalty incurred within a model's representation space as a result of a single amino acid edit at various points along the length of the $\alpha$/$\beta$ CDR3 loops.
To do this, we randomly sampled real TCRs from the testing partition of the Tanno et al. dataset~\cite{tanno_determinants_2020} and synthetically introduced single residue edits in one of their CDR3 loops.
Then, we measured the distance between the original TCR and the single edit variant according to a PLM.
For each model, we sampled TCRs until we had observed at least 100 cases of: 1) each type of edit (insertions, deletions, substitutions) at each position, and 2) substitutions from each amino acid to every other.
Since CDR3 sequences vary in length, we categorised the edit locations into one of five bins: \texttt{C-TERM} for edits within the first one-fifth of the CDR3 sequence counting from the C-terminus, then \texttt{M1}, \texttt{M2}, \texttt{M3}, and \texttt{N-TERM}, in that order.
For this analysis, we investigated SCEPTR and TCR-BERT, since they are the two best performers out of the PLMs tested (Fig.~\ref{fig:benchmarking_overview}d).

\begin{figure*}
  \includegraphics{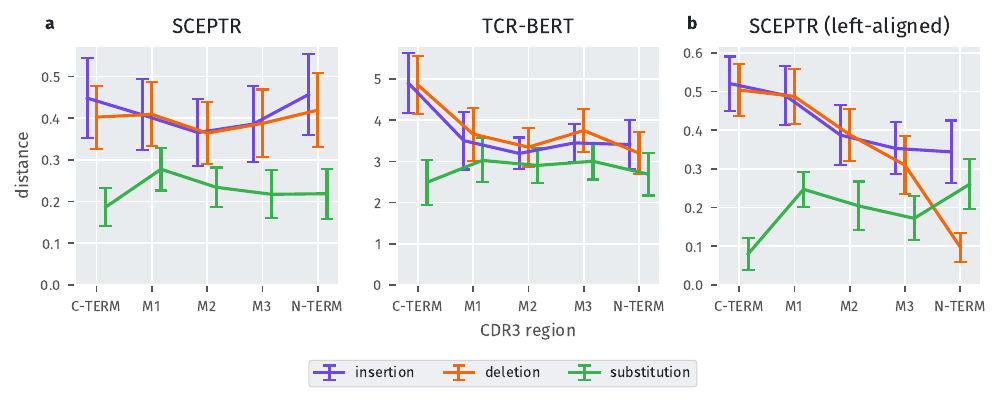}
  \caption{
    \textbf{Investigating TCR co-specificity rules as learned by different PLMs.}
    Here we investigate TCR co-specificity rules as learned by various representation models by measuring the expected distance penalties inucrred by single residue edits in different regions of the $\alpha$ and $\beta$ CDR3 loops.
    We investigate three models: SCEPTR, TCR-BERT, and a SCEPTR variant which replaces its simplified initial embedding module with one that emulates the traditional transformer architecture, including a left-aligned position embedding system (see appendix~\ref{sec:relative_position_embedding}).
    The x axis shows different regions of the CDR3 divided into five bins, where \texttt{C-TERM} represents the first fifth of the loop counting from the C-terminal, \texttt{N-TERM} represents the last fifth of the loop on the N-terminal end, and the middle regions numbered from the C-terminal as shown.
    The y axis hows the expected distance penalty incurred by different types of single edits.
    The different lines show the expected penalty curves with respect to insertions (purple), deletions (orange) and substitutions (green).
    The error bars show the standard deviations.
    According to all models, substitutions on average incur a smaller distance penalty compared to indels.
    While SCEPTR uniformly penalises indels, both TCR-BERT and the left-aligned SCEPTR variant assign higher distance penalties to indels closer to the C-terminal.
  }
  \label{fig:calibration}
\end{figure*}

Both SCEPTR and TCR-BERT generally associate insertions and deletions (indels) with a higher distance penalty compared to substitutions (Fig.~\ref{fig:calibration}a).
While SCEPTR uniformly penalises indels across the length of the CDR3, TCR-BERT assigns higher penalties to those closer to the C-terminus.
We hypothesised that the variation in TCR-BERT's indel penalties is a side-effect of its position embedding system.
TCR-BERT, like many other transformers, encodes a token's position into its initial embedding in a left-aligned manner using a stack of sinusoidal functions with varying periods~\cite{wu_tcr-bert_2021, vaswani_attention_2017, devlin_bert_2019}
This results in embeddings that are more sensitive to indels near the C-terminus, which cause a frame-shift in a larger portion of the CDR3 loop and thus lead to a larger change in the model's underlying TCR representation.
To test this hypothesis, we trained and evaluated a new SCEPTR variant:

\begin{description}
  \item[SCEPTR (left-aligned)] This variant uses a traditional transformer embedding system with trainable token representations and left aligned, stacked sinusoidal position embeddings.
\end{description}

\begin{figure}
  \includegraphics{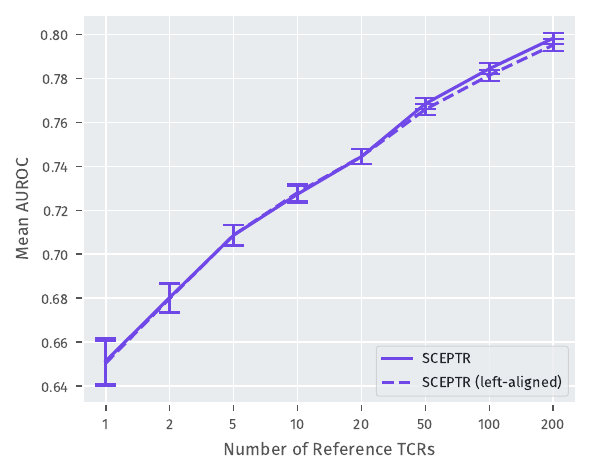}
  \caption{
    \textbf{SCEPTR with its simplified embedder module performs similarly to a variant with an embedder module emulating the traditional transformer architecture.}
    Here we show the results of benchmarking SCEPTR againt a variant which replaces SCEPTR's simplified embedder module (see methods~\ref{sec:methods_sceptr_architecture}) with an implementation emulating the traditional transformer architecture (``left-aligned'' variant in plot, see appendix~\ref{sec:relative_position_embedding}).
    The benchmarking framework as outlined in section~\ref{sec:benchmarking} is used.
    The number of reference sequences varies along the x axis.
    The y axis shows the models' AUROCs averaged across pMHCs.
    We detect no significant difference in performance between the two models.
  }
  \label{fig:simple_vs_classic_embedding}
\end{figure}

While we detect no significant difference in downstream performance between SCEPTR and its left-aligned variant (Fig.~\ref{fig:simple_vs_classic_embedding}), this may be because cases where the differences in their learned rule sets affects performance are rarely seen in our benchmarking data.
The edit penalty profile of the left-aligned variant shows a similar falloff of indel penalties than TCR-BERT with higher penalties at the C than N-terminals (Fig.~\ref{fig:calibration}b).
As their is no clear biological rationale for this observation, these results suggest that SCEPTR's relative position encoding might result in a better-calibrated co-specificity ruleset. These preliminary findings add to the ongoing discussion around how to best encode residue position information in the protein language modelling domain~\cite{elnaggar_ankh_2023}.

Interestingly, the penalty falloff seen with SCEPTR (left-aligned) is sharper than that of TCR-BERT, whose indel penalties plateau past \texttt{M1}.
As TCR-BERT is a substantially deeper model (12 self-attention layers, 12 heads each, embedding dimensionality 768), it might be partially able to internally un-learn the left-aligned-ness of the position information.
If this is true, then position embedding choices are particularly important for training smaller, more efficient models.

\clearpage

\begin{figure}
  \includegraphics{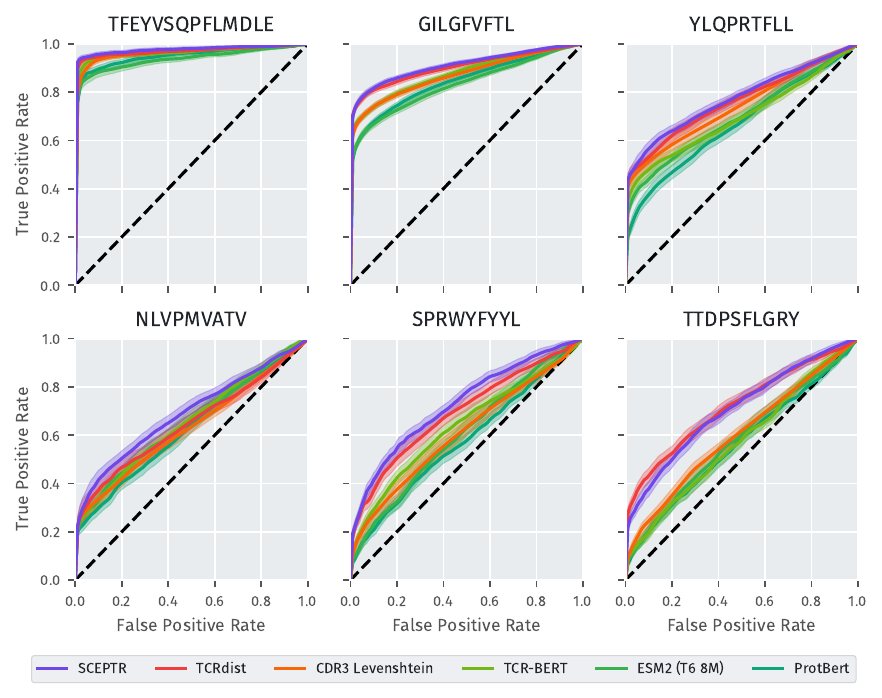}
  \caption{
    \textbf{ROC plots for individual pMHCs on the benchmarking task.}
    Each curve shows the nearest neighbour prediction ROC for a specific model and pMHC for $k=200$ reference TCRs.
    The solid lines correspond to the mean ROC per model per pMHC, and the shaded regions correspond to the standard deviations of the true positive rate across data splits.
    Corresponding per-pMHC AUROC values are provided in table~\ref{tab:per_epitope_aurocs_200_shot}.
  }
  \label{fig:individual_rocs}
\end{figure}

\begin{figure}
  \centering
  \includegraphics{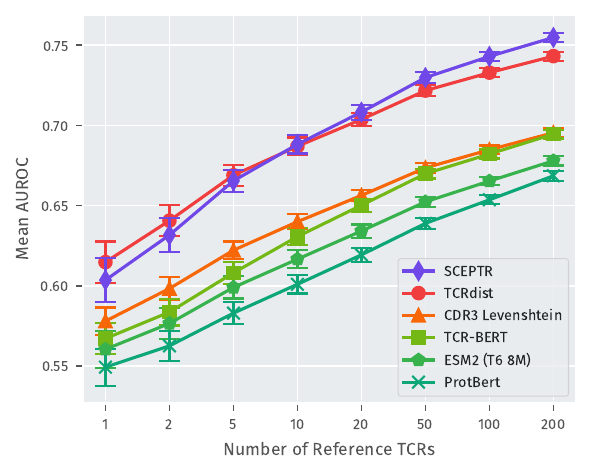}
  \caption{
    \textbf{Benchmarking PLM embeddings on TCR specificity prediction with Montemurro et al.'s post-processed 10xGenomics dataset included.}
    This is a repeat of the benchmarking study from section~\ref{sec:benchmarking} with a larger dataset of labelled TCRs.
    This includes six more sufficiently sampled pMHC specificities (with epitopes \texttt{ELAGIGILTV}, \texttt{GLCTLVAML}, \texttt{AVFDRKSDAK}, \texttt{IVTDFSVIK}, \texttt{RAKFKQLL}, \texttt{KLGGALQAK}).
    The trends seen in section~\ref{sec:benchmarking} are recapitulated.
  }
  \label{fig:benchmarking_with_10x}
\end{figure}

\begin{figure}
  \includegraphics{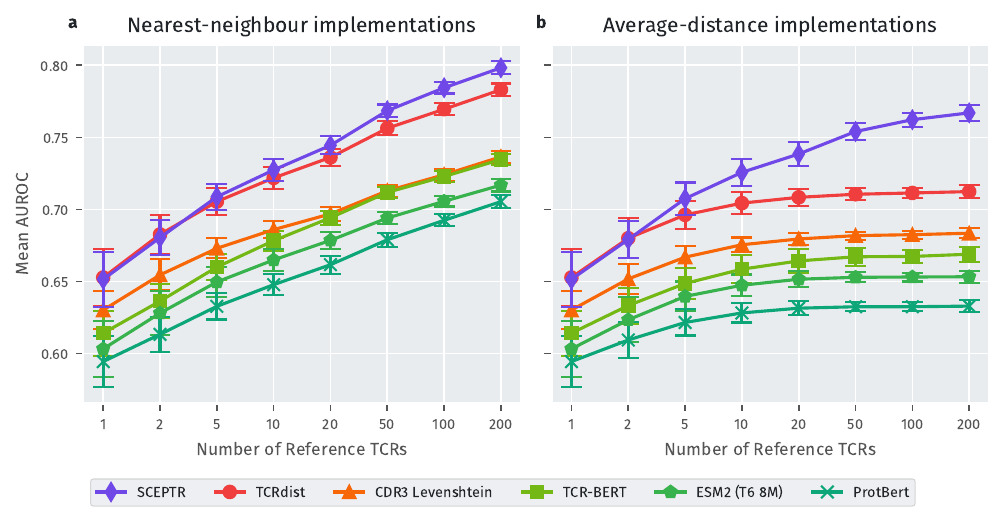}
  \caption{
    \textbf{Nearest neighbour prediction is more performant than using the average distance to all references.}
    We repeated the benchmarking procedure used to produce Fig.~\ref{fig:benchmarking_overview}d, but instead of making inferences based on the distance between a query TCR and its closest reference neighbour we averaged its distance to all references.
    The results are shown in panel \textbf{b)}, with panel \textbf{a)} showing the original nearest neighbour benchmarking results for comparison.
    All models perform better when applied through nearest neighbour prediction.
    Interestingly, when using average distance prediction all models other than SCEPTR rapidly plateau with increasing reference set size.
  }
  \label{fig:benchmarking_avg_dist}
\end{figure}

\begin{figure}
  \includegraphics{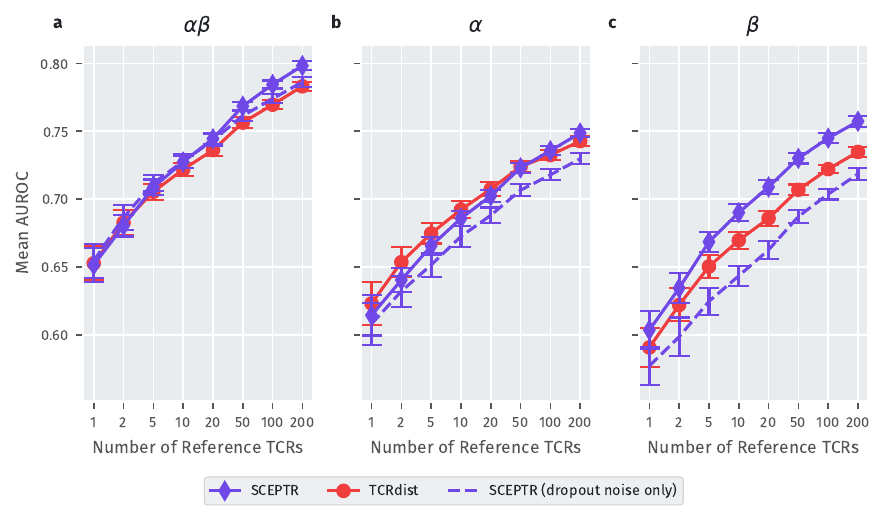}
  \caption{
    \textbf{SCEPTR provides competitive performance on single-chain TCR data.}
    Benchmarking results on \textbf{a)} paired chain data (as in Fig.~\ref{fig:benchmarking_overview}d), or when supplying information about only the \textbf{b)} $\alpha$ or \textbf{c)} $\beta$ chain.
    In each instance, we compared three models: SCEPTR, TCRdist, and a variant of SCEPTR which does not employ any extra noising operations when producing the two views of the same TCR during autocontrastive learning, solely relying on SCEPTR's internal dropout noise (see section~\ref{sec:methods_sceptr_pretraining}).
    In all scenarios, SCEPTR's performance is on par ($\alpha$) or better ($\alpha\beta$ / $\beta$) than TCRdist.
    Furthermore, comparing SCEPTR's performance with that of its dropout noise only variant demonstrates that residue- and chain- dropping during pre-training improves downstream performance, particularly when applied to single-chain data.
  }
  \label{fig:ablation_chain}
\end{figure}

\begin{figure}
    \includegraphics{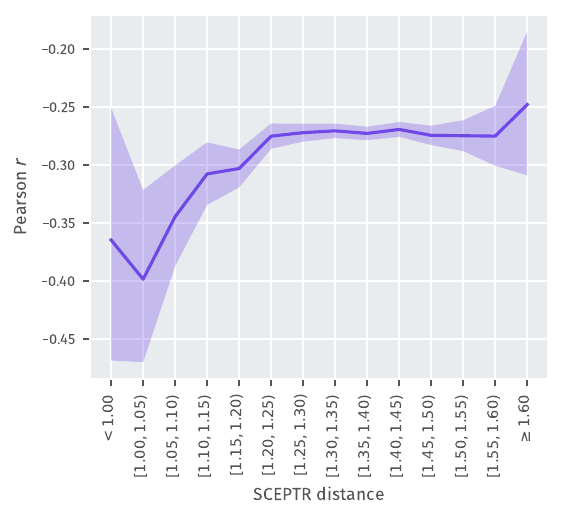}
    \caption{
        \textbf{TCRdist and recombination probabilities are negatively correlated when conditioning on different SCEPTR bins.}
        Conditioned on a certain level of SCEPTR similarity, pairs of sequences with a high probability of recombination ($\p{gen}$) tend to have lower TCRdist distance.
The plot shows the Pearson correlation coefficient $r$ between TCRdist distances and TCR $\p{gen}$ for TCRs across SCEPTR distance bin. The shaded region displays 95\% confidence intervals around estimated correlation coefficients.
    This is a companion to Fig.~\ref{fig:sceptr_vs_tcrdist}, showing the generality of the negative correlation across SCEPTR bins.
    }
    \label{fig:pgen_vs_tcrdist}
\end{figure}

\begin{figure}
  \includegraphics{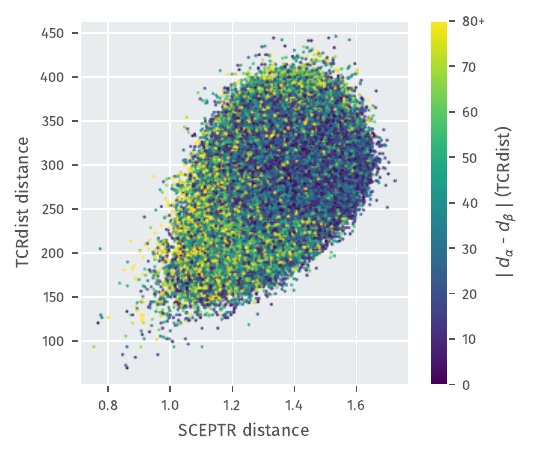}
  \caption{
    \textbf{Large discordance between $\boldsymbol{\alpha}$ and $\boldsymbol{\beta}$ chain similarity explain some of the variation between TCRdist and SCEPTR distances.}
    This plot corresponds to Fig.~\ref{fig:sceptr_vs_tcrdist}a/b in the main text, but points are coloured according to the absolute difference between the $\alpha$ chain and $\beta$ chain components of the TCRdist distance.
    Among TCR pairs with similar overall TCRdist similarity, those pairs which have a large difference between the $\alpha$ and $\beta$ chain TCRdist distances (i.e. where TCRs have one highly similar and another highly dissimilar chain) tend to be assigned lower SCEPTR distances.
  }
  \label{fig:sceptr_vs_tcrdist_chain_delta}
\end{figure}

\begin{figure}
  \includegraphics{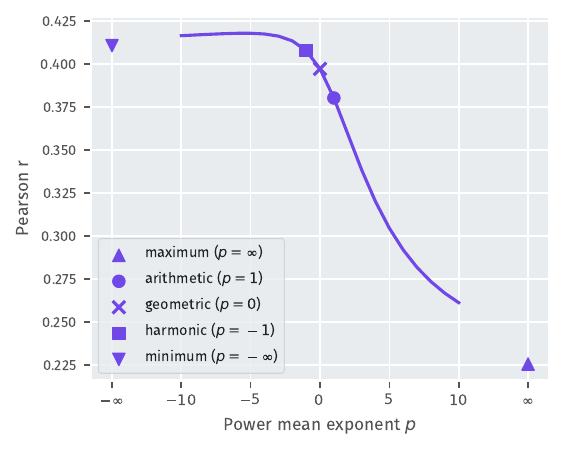}
  \caption{
    \textbf{SCEPTR distances do not average $\boldsymbol{\alpha}$ and $\boldsymbol{\beta}$ chain similarity arithmetically.}
    Fig.~\ref{fig:sceptr_vs_tcrdist_chain_delta} suggests that similarity on at least one chain is sufficient for a low SCEPTR distance.
    To test this hypotheiss, we calculated the Pearson correlation coefficient $r$ between SCEPTR distances and different ways of taking the mean of the $\alpha$ and $\beta$ TCRdist components.
    We interpolated between taking the minimum, arithmetic average, and maximum between the $\alpha$ and $\beta$ chain distances, by computing the generalised power mean of the two distances. The generalised power mean of a set of numbers $x_1, ..., x_n$ is defined as $M_p = \left( \frac{1}{n} \sum_{i=1}^n x_i \right)^{\frac{1}{p}}$, with exponent $p=1$ corresponding to the arithmetic mean, $p=-\infty$ corresponding to taking the minimum and $p=\infty$ corresponding to taking the maximum.
    SCEPTR distances best correlate to power means with exponents $p<1$, suggesting that SCEPTR embeddings behave more like taking the minimum between the chain components ($p=-\infty$), as opposed to the arithmetic average ($p=1$).
    In contrast, the paired-chain TCRdist is defined as the arithmetic average of the distances of both chains.
  }
  \label{fig:sceptr_and_tcrdist_averaging.pdf}
\end{figure}

\begin{figure}
  \includegraphics{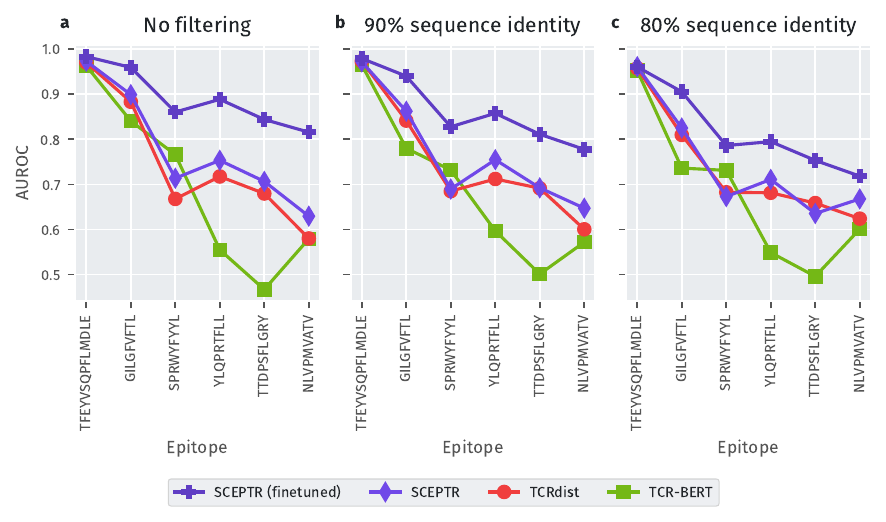}
  \caption{
    \textbf{Supervised contrastive learning improves performance also when filtering highly similar TCRs from the test set.}
    Companion to Fig.~\ref{fig:scl_benchmark} showing the effects of filtering public TCR sequences from the test data, which have exact or near-exact matches in amino acid sequence to any training TCR.
    Benchmarking on \textbf{a)} all data, \textbf{b)} excluding sequences with $\geq$ 90\% sequence similarity, or \textbf{c)} $\geq$ 80\% sequence similarity shows that fine-tuning using supervised contrastive learning goes beyond the memorization of highly similar public TCR motifs.
    Similar sequences were identified using a threshold level of $\alpha$ and $\beta$ CDR3 amino acid sequence identity, and required matching V genes for both chains.
    CDR3 sequence identity was quantified as $1 - (d_\alpha + d_\beta) / (\ell_\alpha + \ell_\beta)$ where $d$ is the Levenshtein edit distance, $\ell$ is the sequence length of the test set TCR's CDR3, and the subscripts denote the two chains.
  }
  \label{fig:scl_discrimination_filtering}
\end{figure}

\begin{figure}
  \includegraphics{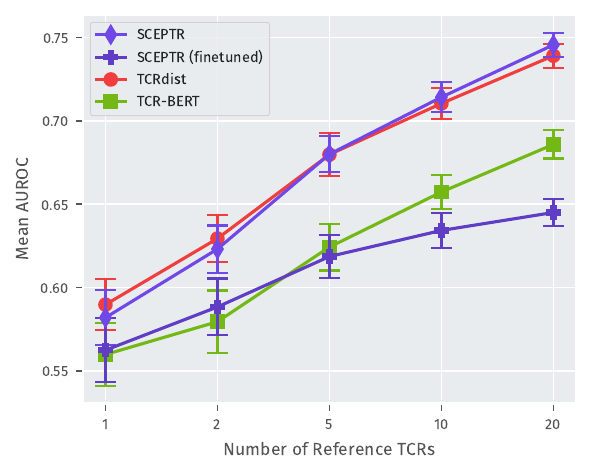}
  \caption{
    \textbf{Benchmarking fine-tuned SCEPTR on TCR specificity prediction for unseen pMHCs.}
    Results of benchmarking fine-tuned SCEPTR (see section~\ref{sec:scl_finetuning}) against TCRdist, TCR-BERT and the baseline SCEPTR model on pMHC targets unseen during SCEPTR's fine-tuning.
    We again use the nearest neighbour benchmarking framework from section~\ref{sec:benchmarking}, but now apply it to pMHCs with more than 120 binders, excluding the six pMHCs used for fine-tuning.
    We restrict the number $k$ of reference sequences to $k \in [1,20]$, to retain at least 100 positive sequences for the calculation of the ROC curves for each data split.
    We find that fine-tuned SCEPTR performs significantly worse compared to the baseline model.
  }
  \label{fig:scl_unseen}
\end{figure}

\begin{figure}
    \includegraphics{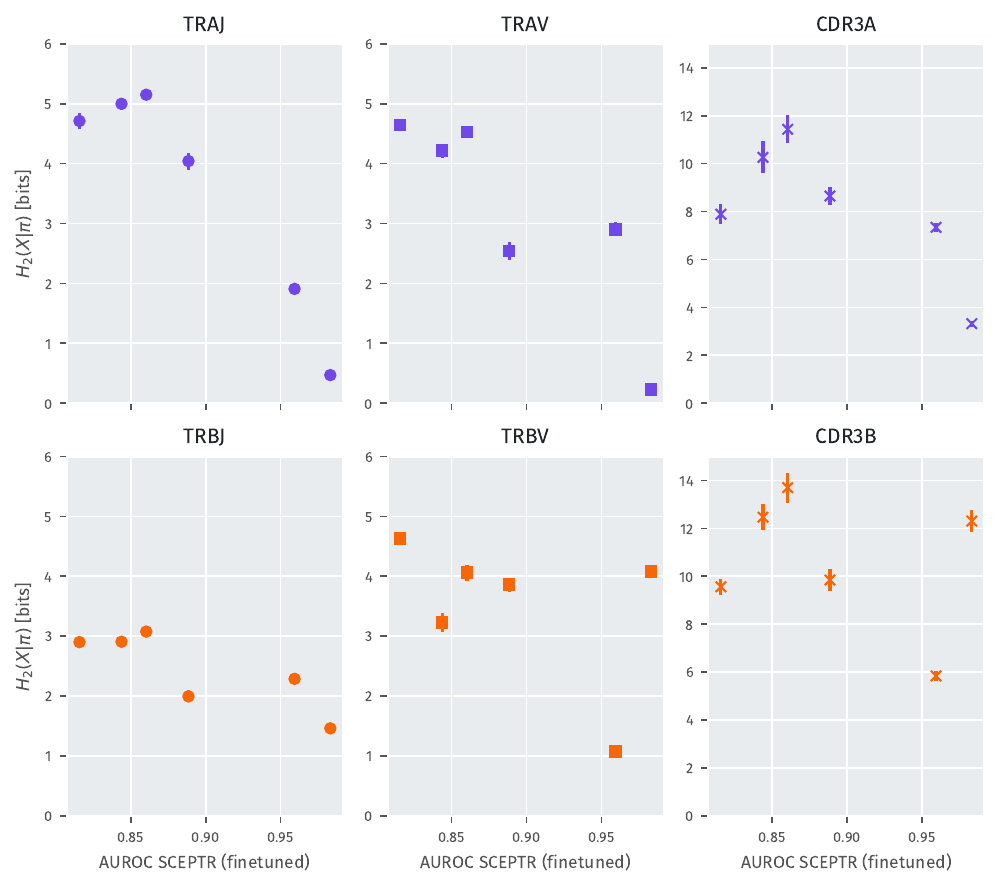}
    \caption{
        \textbf{Sequence diversity of pMHC-specific TCRs used in benchmarking.}
        To investigate why prediction performance of fine-tuned SCEPTR varies across pMHCs, we estimated the coincidence (second-order R\'enyi) entropies~\cite{henderson_limits_2024} of a selection of TCR features (TRAV, TRBV, TRAJ, and TRBJ gene usages, and the amino acid sequences of the $\alpha$ and $\beta$ chain CDR3 loops) among the TCRs specific to each of the six pMHCs used during fine-tuning.
        Each panel displays the feature entropy of TCRs specific to a given pMHC against the fine-tuned SCEPTR variant's AUROC score for the same pMHC.
        The error bars show the standard deviations of the empirical estimates of the coincidence entropies, which were calculated using the unbiased variance estimator for Simpson's diversity described in Ref.~\cite{tiffeau2024unbiased}.
        The easiest to predict pMHCs have lower entropy in most features -- particularly stark reductions in V and J gene diversity are observed for the two pMHCs with the highest AUROCs.
    }
    \label{fig:entropy_vs_performance}
\end{figure}

\begin{figure}
  \includegraphics{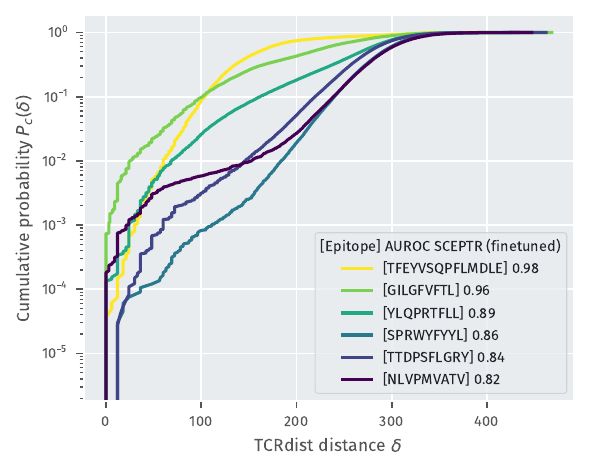}
  \caption{
    \textbf{Statistics of pairwise sequence similarities between pMHC-specific TCRs used in benchmarking.}
    Cumulative probabilities of coincidence are plotted against TCRdist distance threshold for each of the six pMHCs used during SCEPTR fine-tuning.
        The most easy to predict pMHCs have fewer TCR pairs that are highly dissimilar. Interestingly, the TFEYVSQPFLMDLE-specific repertoire, the only class II presented pMHC included in the test set, has the most globally convergent TCRs.
  }
  \label{fig:pc_vs_performance}
\end{figure}

\begin{figure}
  \includegraphics{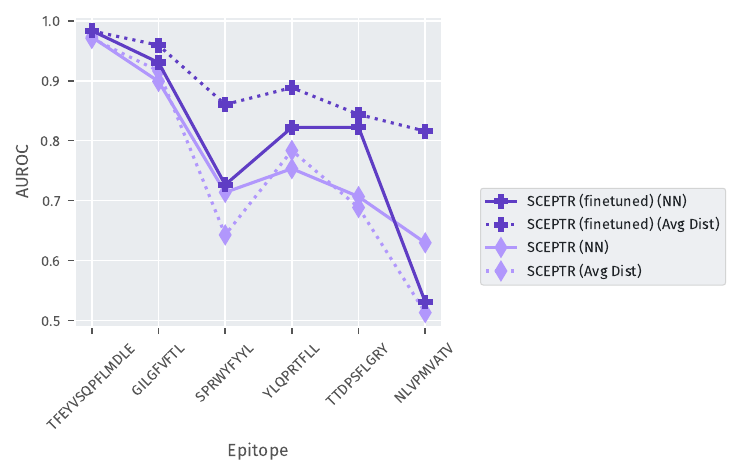}
  \caption{
    \textbf{Comparing nearest neighbour and average distance implementations of the baseline and fine-tuned SCEPTR models.}
    Companion to figure~\ref{fig:scl_benchmark}, comparing the performance of the baseline and fine-tuned versions of SCEPTR using nearest neighbour (NN) or average distance (Avg dist) prediction (see section~\ref{sec:benchmarking}).
    For the baseline model, the nearest neighbour implementation performs better, consistent with the results seen in Fig.~\ref{fig:benchmarking_avg_dist}.
    In contrast, the average distance implementation of the fine-tuned model greatly outperforms its nearest neighbour counterpart.
    Our primary hypothesis as to why most models including the baseline SCEPTR model perform better through nearest neighbour prediction (Fig.~\ref{fig:benchmarking_avg_dist}) is that each pMHC has multiple viable binding solutions comprised of TCRs with different primary sequence features, which means that averaging distance to all reference TCRs across binding solutions dilutes signal.
    The fact that the fine-tuned model no longer shows this property may hint at its ability to better resolve these distinct binding solutions into a single convex cluster.
  }
  \label{fig:sceptr_baseline_finetuned_nn_vs_avg_dist}
\end{figure}

\begin{figure*}
  \centering
  \includegraphics{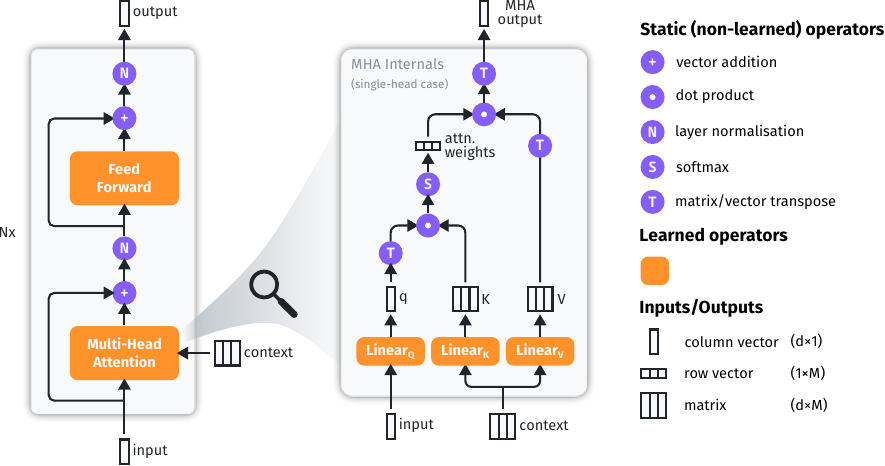}
  \caption{
    \textbf{A simplified schematic depicting the internals of the transformer self-attention stack.}
    The schematic is accurate for the case of a single attention head per layer.
    In the more general case of $H$ attention heads, each MHA block will have $H$ parallel q, k and v linear projections, each from dimensionality $d$ to dimensionality $d/H$.
    Each parallel set of q, k and v vectors/matrices undergo the series of operations shown in the schematic.
    Finally, the final output vector (of shape $d/H \times 1$) from each parallel branch are concatenated together to produce the output of the MHA block (of shape $d \times 1$).
  }
  \label{fig:self_attention_stack}
\end{figure*}

\begin{table}
  \caption{
    Per-epitope summary of the different models' perfomances from the nearest neighbour prediction benchmarking (see section~\ref{sec:benchmarking}) with the number of reference TCRs k=200.
    The best AUROC per epitope is shown in bold.
  }
  \label{tab:per_epitope_aurocs_200_shot}
  \begin{tabular*}{\linewidth}{@{\extracolsep{\fill}} lrrrrrr}
    \toprule
    & SCEPTR & TCRdist & CDR3 Levenshtein & TCR-BERT & ESM2 (T6 8M) & ProtBert \\
    epitope &  &  &  &  &  &  \\
    \midrule
    \texttt{GILGFVFTL} & \textbf{0.911} & 0.904 & 0.872 & 0.876 & 0.831 & 0.845 \\
    \texttt{NLVPMVATV} & \textbf{0.691} & 0.648 & 0.632 & 0.655 & 0.652 & 0.629 \\
    \texttt{SPRWYFYYL} & \textbf{0.728} & 0.695 & 0.610 & 0.637 & 0.604 & 0.575 \\
    \texttt{TFEYVSQPFLMDLE} & \textbf{0.976} & 0.970 & 0.964 & 0.966 & 0.937 & 0.950 \\
    \texttt{TTDPSFLGRY} & 0.708 & \textbf{0.720} & 0.600 & 0.576 & 0.579 & 0.564 \\
    \texttt{YLQPRTFLL} & \textbf{0.775} & 0.762 & 0.743 & 0.698 & 0.697 & 0.669 \\
    \bottomrule
  \end{tabular*}
\end{table}

\begin{table}
  \caption{
    The different studies that contributed TCR data to the training/validation and test splits for the supervised contrastive learning fine-tuning task.
    For datasets without a PubMed ID the table indicates the VDJdb github issue number corresponding to the dataset inclusion.
  }
  \begin{tabular*}{\linewidth}{@{\extracolsep{\fill}} l p{50mm}  >{\raggedright\arraybackslash}p{80mm}<{} }
    \toprule
    epitope & training/validation & test \\
    \midrule
    \texttt{GILGFVFTL} & PMID:28636592 & PMID:12796775, PMID:18275829, PMID:28250417, PMID:28931605, PMID:7807026, PMID:28423320, PMID:28636589, PMID:27645996, PMID:29483513, PMID:29997621, PMID:34793243, \href{https://github.com/antigenomics/vdjdb-db/issues/215}{VDJdbID:215} \\
    \texttt{NLVPMVATV} & PMID:28636592, \href{https://github.com/antigenomics/vdjdb-db/issues/332}{VDJdbID:332} & PMID:19542454, PMID:26429912, PMID:19864595, PMID:28423320, PMID:16237109, PMID:28636589, PMID:36711524, PMID:28623251, PMID:9971792, PMID:17709536, PMID:28934479, PMID:34793243, \href{https://github.com/antigenomics/vdjdb-db/issues/252}{VDJdbID:252} \\
    \texttt{SPRWYFYYL} & PMID:33951417, PMID:35750048 & PMID:33945786, PMID:34793243 \\
    \texttt{TFEYVSQPFLMDLE} & PMID:35750048 & PMID:37030296 \\
    \texttt{TTDPSFLGRY} & PMID:35383307 & PMID:35750048 \\
    \texttt{YLQPRTFLL} & PMID:35383307, PMID:34793243 & PMID:34685626, PMID:37030296, PMID:33664060, PMID:33951417, PMID:35750048, \href{https://github.com/antigenomics/vdjdb-db/issues/215}{VDJdbID:215} \\
    \bottomrule
  \end{tabular*}
  \label{tab:data_split_studies}
\end{table}

\end{document}